\documentclass[journal]{IEEEtran}

\usepackage{cite}
\usepackage{textcomp}
\usepackage{amsmath, amssymb, amsfonts}
\usepackage{accents}
\usepackage{color}
\usepackage{graphicx}
\usepackage{enumerate}
\usepackage{booktabs}
\usepackage{subcaption}
\usepackage{mathrsfs}
\usepackage{mathtools}
\usepackage{dsfont}
\usepackage{relsize}
\usepackage{cmtt}

\usepackage{algorithm}
\usepackage{algpseudocode}

\DeclareMathOperator{\tr}{tr}
\DeclareMathOperator{\minimize}{minimize}

\DeclareMathOperator{\diag}{diag}

\DeclareMathOperator{\E}{\mathsf{E}}
\DeclareMathOperator{\Cov}{\mathsf{cov}}

\DeclareMathOperator{\ProbM}{\mathsf{P}}
\DeclareMathOperator{\Prob}{\mathsf{p}}

\newtheorem{definition}{Definition}
\newtheorem{lemma}{Lemma}

\newtheorem{theorem}{Theorem}
\newtheorem{proposition}{Proposition}
\newtheorem{corollary}{Corollary}
\newtheorem{remark}{Remark}

\newtheorem{assumption}{Assumption}

\begin{document}

\title{Remote State Estimation over Unreliable Channels with Unreliable Feedback: Strategies and Limits}
\author{
Touraj Soleymani, \emph{IEEE}, \emph{Member}, Mohamad Assaad, \emph{IEEE}, \emph{Senior Member}, John S. Baras, \emph{IEEE, Life Fellow}
\thanks{Touraj Soleymani is with the City St George's Department of Engineering, University of London, London EC1V~0HB, United Kingdom (e-mail: {\tt\small touraj.soleymani@citystgeorges.ac.uk}). Mohamad Assaad is with the Telecommunications Department, CentraleSup\'{e}lec, University of Paris-Saclay, 91190 Gif-sur-Yvette, France (e-mail: {\tt\small mohamad.assaad@centralesupelec.fr}). John S.~Baras is with the Institute for Systems Research, University of Maryland College Park, MD 20742, USA (e-mail: {\tt\small baras@umd.edu}).}%
}

\maketitle

\begin{abstract}
In this article, we establish a comprehensive theoretical framework for remote estimation in a networked system composed of a source that is observed by a sensor, a remote monitor that needs to estimate the state of the source in real time, and a communication channel that connects the source to the monitor. The source is a partially observable dynamical process, and the communication channel is a packet-erasure channel with feedback. We consider a novel communication model that captures implicit information. Our main objective is to identify the optimal strategies and the fundamental performance limits of the underlying system in the sense of a causal tradeoff between the packet rate and the mean square error when both forward and backward channels are unreliable. We characterise an optimal coding policy profile consisting of a scheduling policy for an encoder and an estimation policy for a decoder, collocated with the source and the monitor, respectively. We derive the recursive equations that must be solved online by the encoder and the decoder. In addition, we prove that the value function, originally defined over an expanding information set, admits a lower-dimensional representation depending only on two variables. We discuss the structural properties of the optimal policies, and analyse the computational complexity of an algorithm proposed for their computation. We then examine a range of special cases derived from our main theoretical results. We complement the theoretical results with a numerical analysis, and compare the performance of different remote estimation tasks in various operating regimes.
\end{abstract}

\section{Introduction}
Cyber-physical systems are intricate real-time systems that integrate digital and physical components~\cite{kim2012cyber}. These systems harness computation, communication, and control to optimise efficiency, adaptability, and automation in diverse domains, including smart factories, smart healthcare, and smart transportation. A significant challenge within cyber-physical systems arises from their inherent dynamism, which requires continuous status updating to capture the latest information about the physical environment~\cite{sun2017, sun2019, maatouk2020age, kriouile2021global, voi, voi2, touraj-power, soleymaniCUP, soleymani2024found, soleymani2024consis}. As a step towards addressing this challenge, in this article, we develop a framework for remote state estimation in a networked system\footnote{Throughout our study, a networked system is defined as a composed system operating over a communication network, where components interact via exchanged messages subject to communication constraints.} composed of a source that is observed by a sensor, a remote monitor that needs to estimate the state of the source in real time, and a communication channel that connects the source to the monitor. The source is a partially observable dynamical process, and the communication channel is a packet-erasure channel with feedback. This framework is motivated by cyber-physical systems operating over packet-switching networks, where the quality of decision making tightly depends on the availability of sensory observations transmitted as data packets, while packet loss appears as an inevitable phenomenon due to the real-time nature of the operation. Our main objective is to identify the optimal strategies and the fundamental performance limits of the underlying system in the sense of a causal tradeoff between the packet rate and the mean square error when both forward and backward channels are unreliable, a setting that is crucial for designing robust decision-making architectures in modern communication-limited cyber-physical systems. To that end, we search for an optimal coding policy profile that determines the policies of an encoder\footnote{In our study, the encoder is an entity that decides when and what information to transmit based on sensory information, network conditions, and system dynamics.} and a decoder\footnote{In our study, the decoder is an entity responsible for estimating the state of the source in real time based on received information, network conditions, and system dynamics.}, collocated with the source and the monitor, respectively.

\subsection{Related Work}
It is known that reliable communication close to the capacity limit can be attained with error correction subject to infinite delay or with persistent retransmission based on feedback. In remote estimation, where data is real-time, any delay more than a certain threshold is typically intolerable. Moreover, retransmission of stale information is unfavourable when fresh information can be transmitted with the same success rate instead. For these reasons, remote estimation is often conducted in an unreliable communication regime. Previous research has already recognised the severe effects of packet loss, as a consequence of communication unreliability, on stability of Kalman filtering~\cite{sinopoli, plarre, schenato, shi2010, wu2017, huang2007, you2011, quevedo2013, parseh2014, gupta2009d, dey2014remote}. In a seminal work, Sinopoli~\emph{et~al.}~\cite{sinopoli} studied mean-square stability of Kalman filtering over an independent and identically distributed (i.i.d.) channel, and proved that there exists a critical point for the packet error rate above which the expected estimation error covariance is unbounded. To capture the temporal correlation of wireless channels, Wu~\emph{et~al.}~\cite{wu2017} addressed peak-covariance stability of Kalman filtering over a Gilbert--Elliott channel, and proved that there exists a critical region defined by the recovery and failure rates outside which the expected prediction error covariance is unbounded. In addition, to take into account the time variation of the strengths of wireless channels, Quevedo~\emph{et~al.}~\cite{quevedo2013} investigated mean-square stability of Kalman filtering over a fading channel with correlated gains, and established a sufficient condition that ensures the exponential boundedness of the expected estimation error covariance. Differently, Gupta~\emph{et~al.}~\cite{gupta2009d} investigated the estimation of a Gauss--Markov process over a packet-erasure channel, and showed that transmitting the minimum mean-square-error state estimate at the encoder at each time leads to the maximal information set for the decoder. The authors also obtained a necessary and sufficient condition for the packet-loss probability of an i.i.d.~packet-erasure channel that guarantees the boundedness of the expected estimation error~covariance. Dey~\emph{et~al.}~\cite{dey2014remote} also investigated the estimation of a scalar Gauss--Markov process under additive quantisation noise with and without channel feedback. They showed that, when feedback exists, the optimal strategy among all possible linear encoders corresponds to the transmission of the Kalman-filter innovation, and when feedback does not exist, there is a critical packet loss probability above which it is better to transmit the state rather than the innovation. Note that the above works assume that the outputs of the source are transmitted in a periodic manner.

To determine the performance limits of remote estimation, however, one needs to navigate a tradeoff between communication and estimation costs. Various formulations of such a tradeoff have been explored in the literature~\cite{imer2010, lipsa2011, lipsa2009optimal, nayyar2013, molin2017, chakravorty2016, chak2016loss, chak2017loss, rabi2012, guo2021-IT, guo2021-TAC, sun2019, soleymaninecsys, leong2017, leong2018, witsenhausen1979, walrand1983, borkar2001, yuksel2012, tanaka2016}. In particular, the tradeoff between the packet rate and the mean square error, in which a signal-dependent scheduling policy and an estimation policy need to be found, was considered in~\cite{imer2010, lipsa2011, lipsa2009optimal, nayyar2013, molin2017, chakravorty2016, chak2016loss, chak2017loss, rabi2012, guo2021-IT, guo2021-TAC, sun2019, erasure2023, touraj2024tit}. In this formulation, which is of main interest to our study, Lipsa and Martins~\cite{lipsa2011,lipsa2009optimal} analysed the estimation of a scalar Gauss--Markov process over a lossless channel and an i.i.d. channel with reliable feedback. Using majorisation theory, they proved that the optimal scheduling policy is symmetric threshold-based, while the optimal estimation policy is linear switching. Nayyar~\emph{et al.}\cite{nayyar2013} extended \cite{lipsa2011} to an energy-harvesting sensor, where available energy is governed by an external stochastic process, while primarily focusing on a scalar finite-state Markov source and a lossless channel. They proved that the optimal scheduling policy is energy-dependent threshold-based and the optimal estimation policy is the one using the most recently received source state as the current estimate. They also discussed extensions to an orthogonal multi-dimensional Gauss--Markov source, in which state variables evolve independently. In this case, they showed that the optimal estimation policy becomes linear switching. Molin and Hirche~\cite{molin2017} investigated the estimation of a scalar Markov process with symmetric noise distribution over a lossless channel, and devised an iterative algorithm that produces policies similar to those found in~\cite{lipsa2011}. Chakravorty and Mahajan~\cite{chakravorty2016, chak2016loss, chak2017loss} studied the estimation of a scalar autoregressive Markov process with symmetric noise distribution over a lossless channel, an i.i.d.  channel with reliable feedback, as well as a Gilbert--Elliott channel with reliable feedback, and proved based on renewal theory that the optimal scheduling policy is symmetric threshold-based and the optimal estimation policy is linear switching. Guo and Kostina \cite{guo2021-IT} also addressed the estimation of the scalar Ornstein--Uhlenbeck process over a lossless channel, and based on majorisation theory and real induction proved that the optimal scheduling policy and the optimal estimation policy adhere to the same structures.

Note that, in remote estimation, feedback serves as a mechanism that enables the encoder to gain insight into the decoder's belief regarding the state of the source. Most existing studies on remote estimation rely on the assumption of perfect feedback. However, when feedback is imperfect, the knowledge of the encoder about the decoder's belief becomes distorted. This consequently can cause a degradation in the system performance. Despite its crucial significance, there has been limited research on the effects of unreliable feedback on remote estimation~\cite{nourian2014, nourian2014-1, khina2018t, guo2017attack, li2015fake, li2016game, vilni2024goal}. Nourian~\emph{et~al.}~\cite{nourian2014, nourian2014-1} studied the estimation of a Gauss--Markov process over a lossy channel with erroneous feedback. The goal in~\cite{nourian2014} is to minimise the trace of error variance subject to an energy harvesting constraint, and that in~\cite{nourian2014-1} is to minimise a convex combination of the trace of error variance and the communication energy. The authors of these articles used the notion of information state to transform the problem into a simplified one, and accordingly derived the optimal signal-independent scheduling policy and the optimal estimation policy. Khina~\emph{et~al.}~\cite{khina2018t} studied the sequential coding of a Gauss--Markov process with multiple sensors over an i.i.d. channel with delayed feedback, and characterised the achievable causal rate-distortion region. Guo~\emph{et~al.}~\cite{guo2017attack} and Li~\emph{et~al.}~\cite{li2015fake, li2016game} studied the estimation of a Gauss--Markov process over a lossy channel with compromised feedback, and, for some fixed scheduling policies, derived attack policies that target packet acknowledgement signals. Vilni~\emph{et~al.}~\cite{vilni2024goal} also examined status update of a finite-state Markov source when both forward and feedback channels are unreliable and the sensor is energy-harvesting, and developed policies minimising the long-term time average of a distortion function subject to an energy constraint. The above results show that erroneous, delayed, or compromised feedback in general degrades the system performance. Note that, in contrast to the focus of our study, feedback can also be used to enhance channel reliability, provide channel state information, or signal the receipt of control commands. For the effects of unreliable feedback in these cases, we refer the reader to \cite{draper2006noi, draper2008var, mahajan2008des, martins2008cod, mirghaderi2013ach}, to \cite{ekpenyong2006feed, ekpenyong2007feed, love2008over, aggarwal2010pow}, and to \cite{imer2004opt, blind2009, sinopoli2008opt, garone2010lqg, moayedi2013, moon2016robust, lin2016est, lin2017optimal, lin2018st},~respectively.

\begin{figure*}[t]
\centering
  \includegraphics[width=.9\linewidth]{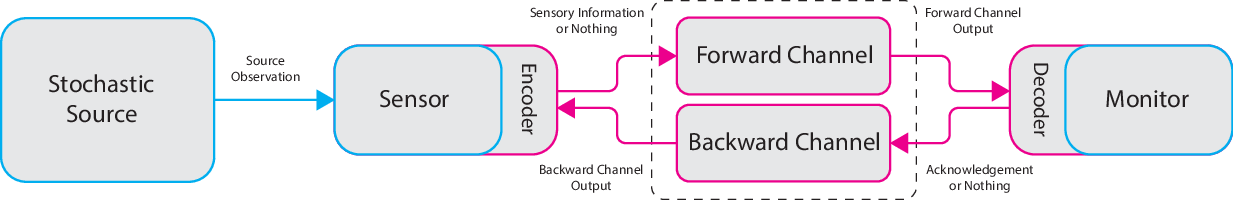}
  \caption{A schematic view of remote state estimation over an unreliable channel with unreliable feedback.}
  \label{fig:block-diagram}
\end{figure*}

\subsection{Contributions and Outline}
In this article, we establish a comprehensive theoretical framework for remote estimation over unreliable channels with unreliable feedback, by considering a novel communication model that captures implicit information. Within the established framework, we characterise a globally optimal coding policy profile consisting of a symmetric threshold-based scheduling policy for the encoder and a linear switching estimation policy for the decoder. We derive the recursive equations that must be solved online by the encoder and the decoder. In addition, we prove that the value function, originally defined over an expanding information set, admits a lower-dimensional representation depending only on two variables. We discuss the structural properties of the optimal policies, and analyse the computational complexity of an algorithm proposed for their computation. We then examine a range of special cases derived from our main theoretical results. We complement the theoretical results with a numerical analysis, and compare the performance of different remote estimation tasks in various operating regimes. A key observation is about the sensitivity of the system performance to the reliability of feedback: we illustrate that the lower the backward error rate, the better the system performance.

We emphasise that our findings contribute broadly to the emerging field of goal-oriented communication~\cite{uysal2022semantic}, which represents a paradigm shift from traditional communication designs toward designs that prioritise the transmission of information based on its relevance to specific tasks or objectives. Unlike traditional approaches that aim to transmit all data as accurately as possible, goal-oriented communication seeks to identify and convey only the most contextually significant portions of data—those that directly impact system performance. In our work, we focus on a key application of this paradigm, i.e., remote estimation in the context of cyber-physical systems. We examine how the importance of sensory data evolves over time and develop mechanisms to assess and prioritise this information sequentially. Through this study, we aim to reduce communication overhead while preserving the effectiveness of remote estimation tasks.

Note that unreliability in the forward and backward channels introduces fundamental challenges in the analysis and design of optimal strategies, which we have rigorously addressed in our paper. Technically speaking, unreliability in the forward channel leads to two forms of implicit information: events related to packet loss and absence of transmission, which provide an opportunity for the monitor to infer additional information beyond what is conveyed through the successful delivery of explicitly transmitted information. Moreover, unreliability in the backward channel compels the encoder to estimate the belief of the decoder, whose task is to estimate the state of the source, at each time. Together, these challenges underscore why designing optimal strategies requires careful structural characterisation.

Our study departs significantly from prior work in several aspects. More specifically, the studies in~\cite{sinopoli, plarre, schenato, shi2010, wu2017, huang2007, you2011, quevedo2013, parseh2014} primarily address stability of remote estimation. In contrast, our focus is on performance optimisation. The studies in~\cite{imer2010, lipsa2011, lipsa2009optimal, nayyar2013, molin2017, chakravorty2016, chak2016loss, chak2017loss, rabi2012, guo2021-IT, guo2021-TAC, sun2019} mainly focus on scalar sources, lossless channels, or channels with lossless feedback. On the contrary, our results hold for multi-dimensional sources and unreliable channels with unreliable feedback. The studies in \cite{vilni2024goal, maatouk2020age, kriouile2021global} focus on finite-state sources, while our work deals with continuous-state sources. Finally, the studies in~\cite{soleymaninecsys, leong2017, leong2018, nourian2014, nourian2014-1, guo2017attack, li2015fake, li2016game} derive variance-based scheduling policies that do not take advantage of realised measurements, and the studies in~\cite{witsenhausen1979, walrand1983, borkar2001, yuksel2012, tanaka2016, khina2018t, gupta2009d, dey2014remote} derive coding policies that operate in a periodic manner. Differently, our results characterise a scheduling policy that depends on realised measurements and operates in a non-periodic manner based on the value of information. 

The article is organised in eight sections. This introductory section is followed by Section~\ref{sec2}, where we formulate the tradeoff problem. We present and discuss our main theoretical results in Sections~\ref{sec:main-results}, \ref{sec:discussion}, and \ref{sec:computation}. We explore several special cases that emerge from our main theoretical results in Section~\ref{sec:specialcases}. We provide our numerical analysis in Section~\ref{sec:example}. Finally, we conclude the article in Section~\ref{sec:conclusion}.

\subsection{Preliminaries}
In the sequel, the sets of real numbers and non-negative integers are denoted by $\mathbb{R}$ and $\mathbb{N}$, respectively. For $x,y \in \mathbb{N}$ and $x \leq y$, the set $\mathbb{N}_{[x,y]}$ denotes $\{z \in \mathbb{N} | x \leq z \leq y\}$. The sequence of all vectors $x_t$, $t=p,\dots,q$, is represented by $\boldsymbol{x}_{p:q}$. For matrices $X$ and $Y$, the relations $X \succ 0$ and $Y \succeq 0$ denote that $X$ and $Y$ are positive definite and positive semi-definite, respectively. The indicator function of a subset $\mathcal{A}$ of a set $\mathcal{X}$ is denoted by $\mathds{1}_\mathcal{A}:\mathcal{X} \to \{0,1\}$. The symbol $\mathcal{O}$ describes the upper bound on the growth rate of a function. The symmetric decreasing rearrangement of a measurable function $f(x)$ vanishing at infinity is represented by $f^*(x)$. The probability measure of a random variable $x$ is represented by $\mathsf{P}(x)$, its probability density or probability mass function by $\Prob(x)$, and its expected value and covariance by $\E[x]$ and $\Cov[x]$, respectively. Stochastic kernels will be adopted to represent decision policies. Let $(\mathcal{X},\mathcal{B}_{\mathcal{X}})$ and $(\mathcal{Y},\mathcal{B}_{\mathcal{Y}})$ be two measurable spaces. A measurable stochastic kernel $\ProbM: \mathcal{B}_{\mathcal{Y}} \times  \mathcal{X} \to [0,1]$ is a mapping such that $\mathcal{A} \mapsto \ProbM( \mathcal{A} | x)$ is a probability measure on $(\mathcal{Y},\mathcal{B}_{\mathcal{Y}})$ for any $x \in \mathcal{X}$, and $x \mapsto \ProbM(\mathcal{A}| x )$ is a measurable function for any $\mathcal{A} \in \mathcal{B}_{\mathcal{Y}}$.

\section{Problem Statement}\label{sec2}
In this section, we mathematically formulate the problem of interest. Our networked system is illustrated in Fig.~\ref{fig:block-diagram}. The state of the source is observed by the sensor. The channel, connecting the sensor to the monitor, is composed of a forward and a backward channel. The encoder, collocated with the source, and the decoder, collocated with the monitor, constitute the two distributed decision makers of the networked system. We first introduce the models of the source and the channel, and then formulate a causal tradeoff between the packet rate and the mean square~error. Throughout this study, we consider a finite time horizon, denoted by~$N$, and assume that the time is discretised into time slots with constant durations.

\subsection{Source Model}
The source is a partially observable Gauss--Markov process, and is described by the discrete-time state and output equations
\begin{align}
	x_{k+1} &= A_k x_k + w_k, \label{eq:sys}\\[1.5\jot]
	y_k &= C_k x_k + v_k, \label{eq:sens}
\end{align}
for $k \in \mathbb{N}_{[0,N]}$ with the initial condition $x_0$, where $x_k \in \mathbb{R}^n$ is the state of the process, $A_k \in \mathbb{R}^{n \times n}$ is the state matrix, $w_k \in \mathbb{R}^n$ is a Gaussian white noise with zero mean and covariance $W_k \succ 0$, $y_k \in \mathbb{R}^m$ is the output of the process, $C_k \in \mathbb{R}^{m \times n}$ is the output matrix, and $v_k \in \mathbb{R}^m$ is a Gaussian white noise with zero mean and covariance $V_k \succ 0$.

\begin{assumption}
For the source model in (\ref{eq:sys}) and (\ref{eq:sens}), the following assumptions are satisfied:
\begin{enumerate}[(i)]
	\item The initial condition $x_0$ is a Gaussian vector with mean $m_0$ and covariance $M_0$.
	\item The random variables $x_0$, $w_t$, and $v_s$ for $t,s \in \mathbb{N}_{[0,N]}$ are mutually independent, i.e., $\Prob(x_0, \boldsymbol{w}_{0:N}, \boldsymbol{v}_{0:N}) = \Prob(x_0) \prod_{k=0}^{N} \Prob(w_k) \prod_{k=0}^{N} \Prob(v_k)$.
\end{enumerate}
\end{assumption}

\begin{remark}
The source model given in (\ref{eq:sys}) and (\ref{eq:sens}) represents a wide class of dynamical processes. Note that such a Gauss--Markov model has been adopted extensively in the fields of communication and control, since this model provides a foundation for development of more sophisticated systems. Moreover, the time-varying nature of the model allows approximation of nonlinear systems around their nominal trajectories, and its partially observable nature respects the fact that in reality only a noisy version of the output can be observed. Clearly, a model that is time-invariant and/or fully observable is a special case of this model.
\end{remark}

\subsection{Channel Model}
Both forward and backward channels are packet-erasure channels. At each time $k$, a message containing sensory information, represented by $\check{x}_k$, can be sent by the encoder over the forward channel, which has one-step delay and is i.i.d.~with forward error rate $\lambda$, to the decoder, where an estimate of the state of the process, represented by $\hat{x}_k$, should be computed. Let $u_k \in \{0,1\}$ be a binary variable such that $u_k = 1$ if $\check{x}_k$ is transmitted at time $k$, and $u_k = 0$ otherwise. A transmitted message in the forward channel at each time is successfully received by the decoder with probability $\lambda^c := 1 - \lambda$. Let $\gamma_k \in \{0,1\}$ be a binary random variable modelling the packet loss in the forward channel such that $\gamma_k = 0$ if a packet loss occurs in the forward channel when it is used to exchange $\check{x}_k$, and $\gamma_k = 1$ otherwise. Then, $\Prob(\gamma_k=0) = \lambda$ for $k \in \mathbb{N}_{[0,N]}$. Accordingly, the output of the forward channel at time $k+1$ is specified as
\begin{align}\label{eq:forward-model}
z^f_{k+1} = \left\{
  \begin{array}{l l}
     \check{x}_k, & \ \text{if} \ u_k = 1 \ \wedge \ \gamma_k =1, \\[1\jot]
     \mathfrak{F}, & \ \text{if} \ u_k = 1 \ \wedge \ \gamma_k =0, \\[1\jot]
     \mathfrak{E}, & \ \text{otherwise},
  \end{array} \right.
\end{align}
for $k \in \mathbb{N}_{[0,N]}$ with $z^f_0 = \mathfrak{E}$ by convention, where $\mathfrak{F}$ and $\mathfrak{E}$ represent packet loss and absence of transmission in the forward channel, respectively. At each time $k$, if the forward channel is used to exchange $\check{x}_k$, i.e., if $u_k = 1$, a packet acknowledgement is sent back by the decoder to the encoder over the backward channel, which has zero delay and is i.i.d. with backward error rate $\rho$. A transmitted message in the backward channel at each time is successfully received by the encoder with probability $\rho^c := 1 - \rho$. Let $\theta_k \in \{0,1\}$ be a binary random variable modelling the packet loss in the backward channel such that $\theta_k = 0$ if a packet loss occurs in the backward channel when it is used to exchange $\gamma_k$, and $\theta_k = 1$ otherwise. Then, $\Prob(\theta_k=0) = \rho$ for $k \in \mathbb{N}_{[0,N]}$. Accordingly, the output of the backward channel at time $k+1$ is specified as
\begin{align}\label{eq:backward-model}
z^b_{k+1} = \left\{
  \begin{array}{l l}
     \gamma_{k}, & \ \text{if} \ u_k = 1 \ \wedge \ \theta_{k} = 1, \\[1\jot]
     \mathfrak{H}, & \ \text{if} \ u_k = 1 \ \wedge \ \theta_{k} = 0, \\[1\jot]
     \mathfrak{G}, & \ \text{otherwise},
  \end{array} \right.
\end{align}
for $k \in \mathbb{N}_{[0,N]}$ with $z^b_0 = \mathfrak{G}$ by convention, where $\mathfrak{H}$ and $\mathfrak{G}$ represent packet loss and absence of transmission in the backward channel, respectively.

\begin{assumption}
For the channel model in (\ref{eq:forward-model}) and (\ref{eq:backward-model}), the following assumptions are satisfied:
\begin{enumerate}[(i)]
	\item The random variables $\gamma_t$ and $\theta_s$ for $t,s \in \mathbb{N}_{[0,N]}$ are mutually independent, i.e., $\Prob(\boldsymbol{\gamma}_{0:N}, \boldsymbol{\theta}_{0:N}) = \prod_{k=0}^{N} \Prob(\gamma_k) \prod_{k=0}^{N} \Prob(\theta_k)$. 
	\item By convention, set $\gamma_k =\theta_k = 0$ when nothing is transmitted in the forward channel at time $k$, i.e., $u_k = 0$.
	\item Quantisation error is negligible, i.e., in a successful transmission, real-value sensory information can be conveyed from the encoder to the decoder without any error.
\end{enumerate}
\end{assumption}

\begin{remark}
The channel model given in (\ref{eq:forward-model}) and (\ref{eq:backward-model}) lends itself to the representation of different communication regimes. In particular, we have reliable communication\footnote{Throughout this study, reliable communication and reliable feedback refer to the idealised cases $\lambda = 0$ and $\rho = 0$, respectively. We use these regimes as theoretical benchmarks. In practice, reliability typically implies very low but non-perfect error rates (e.g., $<0.0001$).} without acknowledgements when $\lambda = 0$ and $\rho = 1$, unreliable communication with reliable acknowledgements when $\lambda > 0$ and $\rho = 0$, unreliable communication without acknowledgements when $\lambda > 0$ and $\rho = 1$, and unreliable communication with unreliable acknowledgements when $\lambda > 0$ and $\rho > 0$ (i.e., the general regime). Note that the forward and backward channels are assumed to operate with one-step delay and zero delay, respectively, as the main communication load is in the forward channel. Moreover, the effect of the quantisation error is assumed to be negligible, as the number of bits in each data packet is often sufficiently large in packet-switching networks, which is a common assumption in the literature of remote estimation~\cite{imer2010, lipsa2011, lipsa2009optimal, molin2017, chakravorty2016, chak2016loss, chak2017loss, rabi2012, guo2021-IT, guo2021-TAC, sun2019, soleymaninecsys, leong2017, leong2018}.
\end{remark}

\begin{remark}
Note that according to (\ref{eq:forward-model}), aside from the sensory information that can be encoded explicitly at time $k$ as $\check{x}_k$, implicit information can also be encoded through $\mathfrak{F}$ and $\mathfrak{E}$. To illustrate this, suppose that $u_k = 1$ if and only if $y_k \in \mathcal{A}_k$, where $\mathcal{A}_k$ can be any measurable set. Then, the decoder can infer that $y_k \in \mathcal{A}_k$ or $y_k \notin \mathcal{A}_k$ when $\mathfrak{F}$ or $\mathfrak{E}$ is detected at time $k+1$, respectively. Moreover, note that according to (\ref{eq:backward-model}), $\mathfrak{H}$ is detected by the encoder at time $k+1$ when sensory information is transmitted in the forward channel at~time~$k$ and a packet loss occurs in the backward channel at time $k+1$, and $\mathfrak{G}$ is detected at time $k+1$ when nothing is transmitted in the backward channel at~time~$k$.
\end{remark}

\subsection{A Team Decision-Making Problem}
In our framework, the decision variables $u_k$ and $\hat{x}_k$ at each time $k$ for $k \in \mathbb{N}_{[0,N]}$ are determined by the encoder and the decoder, respectively, as a team in a distributed way. This distributed decision-making process relies on causal information. Let the information sets of the encoder and the decoder be denoted by $\mathcal{I}^e_k$ and $\mathcal{I}^d_k$, which, in general, are specified~as
\begin{align}
	\mathcal{I}^e_k &= \Big\{ y_t, z^b_t, u_s, \theta_s \big| t \in \mathbb{N}_{[0,k]}, s \in \mathbb{N}_{[0,k-1]} \Big\}, \label{eq:infoset1}\\[1\jot]
	\mathcal{I}^d_k &= \Big\{ z^f_t, u_s, \gamma_s \big| t \in \mathbb{N}_{[0,k]}, s \in \mathbb{N}_{[0,k-1]} \Big\}. \label{eq:infoset2}
\end{align}

A coding policy profile $(\epsilon,\delta)$ consisting of a scheduling policy $\epsilon$ and an estimation policy $\delta$ is admissible if $\epsilon = \{ \ProbM(u_k | \mathcal{I}^e_k) \}_{k=0}^{N}$ and $\delta = \{ \ProbM(\hat{x}_k | \mathcal{I}^d_k) \}_{k=0}^{N}$, where $\ProbM(u_k | \mathcal{I}^e_k)$ and $\ProbM(\hat{x}_k | \mathcal{I}^d_k)$ are measurable stochastic kernels. In the sequel, we seek to find a globally optimal solution $(\epsilon^\star,\delta^\star)$ to a team decision-making problem cast as
\begin{align}\label{eq:main_problem1}
	&\underset{\epsilon \in \mathcal{E},\delta\in \mathcal{D}}{\minimize} \ \Phi(\epsilon,\delta),
\end{align}
subject to the source model in (\ref{eq:sys}) and (\ref{eq:sens}), and the channel model in (\ref{eq:forward-model}) and (\ref{eq:backward-model}), where $\mathcal{E}$ and $\mathcal{D}$ are the sets of admissible scheduling policies and admissible estimation policies, and 
\begin{align}\label{eq:loss-function}
\Phi(\epsilon,\delta) := \E \bigg[ \sum_{k=0}^{N} \alpha_k u_k + ( x_k - \hat{x}_k)^T ( x_k - \hat{x}_k) \bigg],
\end{align}
for the weighting coefficient $\alpha_k \geq 0$, which specifies communication cost per transmission. The concept of global optimality in the context of our team decision-making problem is captured by the following definition.

\begin{definition}[Global optimality]
A policy profile $(\epsilon^\star,\delta^\star)$ in the above team decision-making problem with two distributed decision makers is globally optimal if
\begin{align*}
	\Phi(\epsilon^\star,\delta^\star) \leq \Phi(\epsilon,\delta), \ \text{for all } \epsilon \in \mathcal{E}, \delta \in \mathcal{D}.
\end{align*}
\end{definition}

\begin{remark}
The loss function in (\ref{eq:loss-function}) is expressed in terms of two performance criteria: mean square error and packet rate. Note that we omitted the coefficient $\frac{1}{N+1}$ in these criteria, as it does not affect the solutions to the problem in~(\ref{eq:main_problem1}). The mean-square-error criterion, which is frequently used in the analysis of control systems, represents the quality of estimation. However, the packet-rate criterion, which is frequently used in the analysis of packet-switching networks, represents the cost of communication. Note that, apart from the time averaging, the expectation is adopted in the definition of the loss function in order to formulate an optimisation problem whose solution is not path-dependent (see e.g., \cite{bertsekas1995DP}). We should also emphasise that the problem in~(\ref{eq:main_problem1}) is a decision-making problem with a non-classical information structure. This problem is nonconvex and in general intractable. Finding a globally optimal solution $(\epsilon^\star,\delta^\star)$ to this problem enables us to determine the fundamental performance limits of the networked~system.
\end{remark}

The next definition introduces a value function related to the loss function in (\ref{eq:loss-function}), which will be instrumental in our analysis.
\begin{definition}[Value function]
The value function $V_k(\mathcal{I}^e_k)$ associated with the loss function $\Phi(\epsilon,\delta)$ and the information set $\mathcal{I}^e_k$ is defined by
\begin{align}
	V_k(\mathcal{I}^e_k) :=& \min_{\epsilon \in \mathcal{E} : \delta = \delta^\star}\E \bigg[ \sum_{t=k}^{N-1} \alpha_t u_t + \hat{e}_{t+1}^T \hat{e}_{t+1} \Big| \mathcal{I}^e_k \bigg], \label{eq:Ve-def}
\end{align}
for $k \in \mathbb{N}_{[0,N]}$, where the estimation policy $\delta^\star$ is chosen based on $\hat{x}_k = \E[x_k | \mathcal{I}^d_k]$.
\end{definition}

\begin{remark}
The value function $V_k(\mathcal{I}^e_k)$, defined over the space of the encoder's information set $\mathcal{I}^e_k$, represents the minimum cumulative loss incurred from time $k$ to the terminal time $N$, under the assumption that the estimation policy $\delta^\star$ is fixed and given by $\hat{x}_k = \E[x_k \mid \mathcal{I}^d_k]$. This value function is a quantity from the perspective of the encoder, and incorporates the impact of the optimal scheduling decisions. Note that, due to causality, the scheduling policy $\epsilon \in \mathcal{E}$ at each time $k$ can affect $\hat{e}_t$ only from time $t=k+1$ onward. This implies that at the final time step $k = N$, the encoder has no further opportunity to influence the estimation process through its scheduling decision. Hence, $u_N$ should be equal to zero in the optimal scheduling policy~$\epsilon^\star$.
\end{remark}

\section{Main Results}\label{sec:main-results}
In this section, we present the main theoretical results on the synthesis of a globally optimal policy profile. Let $\check{x}_k = \E[x_k | \mathcal{I}^e_k]$ and $\hat{x}_k = \E[x_k | \mathcal{I}^d_k]$, and define the innovation from the perspective of the encoder $\nu_k := y_k - C_k \E [x_k | \mathcal{I}^e_{k-1}]$, the estimation error from the perspective of the decoder $\hat{e}_k := x_k - \E[x_k | \mathcal{I}^d_k]$, the estimation mismatch $\tilde{e}_k := \E[x_k | \mathcal{I}^e_k] - \E[x_k | \mathcal{I}^d_k]$, the conditional mean of the estimation mismatch from the perspective of the encoder $\breve{e}_k := \E[ \tilde{e}_k | \mathcal{I}^e_k]$, the conditional covariance of the estimation mismatch from the perspective of the encoder $R_k := \Cov[ \tilde{e}_k | \mathcal{I}^e_k]$, and the value function residual $\Delta_k := \E[ V_{k+1}(\mathcal{I}^e_{k+1}) | \mathcal{I}^e_k, u_k = 0] - \E[ V_{k+1}(\mathcal{I}^e_{k+1}) | \mathcal{I}^e_k, u_k = 1]$. 

The following theorem provides a globally optimal policy profile for the problem in (\ref{eq:main_problem1}).

\begin{theorem}\label{thm:1}\emph{
The causal tradeoff between the packet rate and the mean square error under unreliable communication with unreliable acknowledgements admits a globally optimal solution $(\epsilon^\star, \delta^\star)$ such that $\epsilon^\star$ is a symmetric threshold-based scheduling policy given~by
\begin{align}
	u_k = \mathds{1}_{\chi_k - \alpha_k \geq 0},
\end{align}
for $k \in \mathbb{N}_{[0,N-1]}$, in conjunction with $\check{x}_k = \E[x_k | \mathcal{I}^e_k]$, where $\chi_k(\breve{e}_k, R_k) = \lambda^c \breve{e}_k^T A_k^T A_k \breve{e}_k + \lambda^c \tr(A_k R_k A_k^T) + \Delta_k$ is a symmetric function of $\breve{e}_k$ and a function of $R_k$, which requires solving
\begin{align}
	\check{x}_{k} &= A_{k-1} \check{x}_{k-1} + K_{k} \nu_k, \label{eq:est-KF-xhat:thm}\\[1.75\jot]
	Q_{k} &= \big( (A_{k-1} Q_{k-1} A_{k-1}^T + W_{k-1})^{-1} + L_k \big)^{-1}, \label{eq:KF-cov:thm}\\[2\jot]
	\breve{e}_k &= (1 - \bar{\theta}_{k-1}) A_{k-1} \breve{e}_{k-1} + K_k \nu_k, \label{eq:eps:thm}\\[2\jot]
	R_k &= (1 - \bar{\theta}_{k-1}) A_{k-1} R_{k-1} A_{k-1}^T  + \bar{\bar{\theta}}_{k-1} H_{k-1}, \label{eq:R:thm}	
\end{align}
for $k \in \mathbb{N}_{[1,N]}$ with initial conditions $\check{x}_0 = m_0 + K_0 \nu_0$, $Q_0 = (M_0^{-1} + C_{0}^T V_{0}^{-1} C_{0})^{-1}$, $\breve{e}_0 = K_0 \nu_0$, and $R_0 = 0$, where $\nu_k = y_k - C_k A_{k-1} \check{x}_{k-1}$, $K_{k} = Q_{k} C_{k}^T V_{k}^{-1}$, $L_k = C_{k}^T V_{k}^{-1} C_{k}$, $\bar{\theta}_{k-1} = \theta_{k-1} u_{k-1} \gamma_{k-1} + \lambda^c (1 - \theta_{k-1}) u_{k-1}$, $\bar{\bar{\theta}}_{k-1} = (\lambda^c - {\lambda^c}^2) (1 - \theta_{k-1} )  u_{k-1}$, and $H_{k-1} = A_{k-1} \breve{e}_{k-1} \breve{e}_{k-1}^T A_{k-1}^T$; and $\delta^\star$ is a linear switching estimation policy given by
\begin{align}
	\hat{x}_k = A_{k-1} \hat{x}_{k-1} + u_{k-1} \gamma_{k-1} A_{k-1} \tilde{e}_{k-1}, \label{eq:xh:thm}
\end{align}
for $k \in \mathbb{N}_{[1,N]}$ with initial condition $\hat{x}_0 = m_0$, where $\hat{x}_k = \E[x_k | \mathcal{I}^d_k]$, without being influenced by implicit information.}
\end{theorem}
\begin{IEEEproof}
The proof proceeds as follows. We first establish that, without loss of optimality, the encoder can transmit the conditional mean $\check{x}_k = \mathbb{E}[x_k \,|\, \mathcal{I}^e_k]$ and the decoder can compute the state estimate $\hat{x}_k = \mathbb{E}[x_k \,|\, \mathcal{I}^d_k]$. We then complete the argument in three steps:
\begin{enumerate}[(i)]
	\item In the first step, we show that $\Phi(\epsilon^n,\delta^o) = \Phi(\epsilon^o,\delta^o)$, where $\epsilon^n$ is an innovation-based scheduling policy, i.e., it can be expressed in the form of $\Prob(u_k | \boldsymbol{\nu}_{0:k}, \boldsymbol{z}^b_{0:k}, \boldsymbol{u}_{0:k-1}, \boldsymbol{\theta}_{0:k-1})$.
	\item  In the second step, we show that $\Phi(\epsilon^s,\delta^o) \leq \Phi(\epsilon^n,\delta^o)$, where $\epsilon^s$ is a special form of $\epsilon^n$ that is symmetric with respect to $\boldsymbol{\nu}_{0:k}$.
	\item In the final step, we show that $\Phi(\epsilon^\star,\delta^\star) \leq \Phi(\epsilon^s,\delta^o)$, where $\epsilon^\star$ is a special form of $\epsilon^s$ that satisfies the Bellman optimality equation and $\delta^\star$ is the same as $\delta^o$, constructed based on $\hat{x}_k = \E[x_k | \mathcal{I}^d_k]$.
\end{enumerate}
See the Appendix for the complete proof.
\end{IEEEproof}

\begin{remark}
In the structural results of Theorem~\ref{thm:1}, the scheduling policy $\epsilon^\star$ at each time $k$ depends on $\breve{e}_k$ and $R_k$, which evolve according to the recursive equations (\ref{eq:eps:thm}) and (\ref{eq:R:thm}). The preceding terms in (\ref{eq:eps:thm}) and (\ref{eq:R:thm}) are responsible for the accumulations of the conditional means and the conditional covariances of the estimation mismatches. These terms are completely eliminated only if $u_{k-1} \gamma_{k-1} \theta_{k-1} = 1$, i.e., there is a successful transmission of sensory information in the forward channel followed by a successful transmission of an acknowledgement in the backward channel. Otherwise, the accumulated errors grow persistently without correction. In addition, the estimation policy $\delta^\star$ at each time $k$ is given by $\hat{x}_k$, which satisfies the recursive equation (\ref{eq:xh:thm}). The preceding term in (\ref{eq:xh:thm}) is responsible for the accumulation of the state estimates from the perspective of the decoder. This term is completely eliminated only if $u_{k-1} \gamma_{k-1} = 1$, i.e., there is a successful transmission of sensory information in the forward channel. Otherwise, the estimate evolves purely based on system dynamics without correction. Further discussion about the main results is provided in the following section.
\end{remark}

\section{Discussion}\label{sec:discussion}
In this section, we discuss the salient structural characteristics of the main results. First, we note that the information structure of the networked system, represented by $( \{ \mathcal{I}^e_k \}_{k=0}^{N}, \{ \mathcal{I}^d_k \}_{k=0}^{N})$, is non-classical because any decision of the encoder can change the information available to the decoder, while the decoder does not have access to the information used in the construction of that decision (see e.g., \cite{yuksel} for further discussions about non-classical information structures). This is actually one of the main difficulties in finding an optimal coding policy profile for our team decision-making problem. Furthermore, we note that, based on the information structure and the fact that $\mathcal{I}^d_k \not\subset \mathcal{I}^e_k$, the knowledge of the encoder about the decoder's belief at each time $k$ is imperfect. As a result, the encoder has to estimate the decoder's belief before making any decision, which complicates the structures of the decision~policies.

On one hand, an admissible scheduling policy is generally defined based on a condition like $h_k(\mathcal{I}^e_k) \in \mathcal{H}_k$, where $h_k(.)$ and $\mathcal{H}_k$ are a measurable function and a measurable set, respectively (see e.g., \cite{sijs2012} for more details). As a result, such a policy should typically necessitate complex computation. Our results, however, indicate that there exists a globally optimal scheduling policy $\epsilon^\star$ that is threshold-based. This structure significantly simplifies the design of the encoder, as now one only needs to search for a threshold value at each time~$k$. The results also state that the encoder is required to run four recursive equations. On the other hand, an admissible estimation policy is generally dependent on a conditional distribution of the form $\ProbM(x_k | \mathcal{I}^d_k)$, which is non-Gaussian due to the signalling effect (see e.g., \cite{wu2013} for more details). The signalling effect exists because the decoder can still update its belief even when it does not receive any sensory information, i.e., $u_k = 0 \vee \{u_k = 1 \wedge \gamma_k =0\}$. As a result, such a policy should typically be nonlinear and without any analytical form. Our results, however, indicate that there exists a globally optimal estimation policy $\delta^\star$ that is linear switching without being influenced by the implicit information. This structure dramatically simplifies the design of the decoder. The results also state that the decoder is required to run only one recursive equation, which is independent of the backward channel condition. Together, these results enable tractable computation of the performance limits for the networked system.

We should point out that the decoupled design is valid and optimal when the encoder's scheduling policy is symmetric threshold-based and the decoder's estimation policy is a linear switching filter unaffected by implicit information, as described in Theorem~\ref{thm:1}. These properties hold despite the non-classical information structure, due to the specific dynamical models and cost criteria established in our problem. In general, the decoupled design may not hold. For instance, fixing the scheduling policy a priori to an arbitrary state-dependent policy can lead to an optimal estimation policy that is tightly coupled with the structure of the scheduling policy (see e.g., \cite{imer2010, xu2020remote}). This implies that a Kalman-like filter for the decoder, which is independent of the structure of the scheduling policy, is not always optimal for all~encoders.

\section{Computational Aspects}\label{sec:computation}
In this section, we explore the computational aspects associated with the implementation of the policy profile characterised in Theorem~\ref{thm:1}. According to the main results, the value function $V_k(\mathcal{I}^e_k)$ can be transformed into $V_k(\breve{e}_k, R_k)$, defined over the space of the conditional mean $\breve{e}_k$ and the conditional covariance $R_k$ of the estimation mismatch. These variables evolve in a nonlinear and stochastic manner over time. As $\breve{e}_k$ and $R_k$ are continuous-valued, we can discretise both over appropriately chosen grids with arbitrary resolutions. For each discretised pair $(\breve{e}_k, R_k)$, we can then compute the value function $V_k(\breve{e}_k, R_k)$ by solving the following recursive equation backward in time:
\begin{align}\label{eq:compute:V}
V_{k}(\breve{e}_k, R_k) = \min \Big\{  V_k^{(0)}, V_k^{(1)} \Big\},
\end{align}
where
\begin{align}\label{eq:Vku}
&V_k^{(u_k)} = \alpha_k u_k + (1 - \lambda^c u_k) \breve{e}_k^T A_k^T A_k \breve{e}_k \nonumber\\[2\jot]
	&\qquad + (1 - \lambda^c u_k) \tr(A_k R_k A_k^T) \nonumber\\[1.5\jot]
	&\qquad + \tr \Big( Q_{k+1} C_{k+1}^T V_{k+1}^{-1} \nonumber\\[1.5\jot]
	&\qquad \quad \qquad \times \big(C_{k+1} (A_k Q_k A_k^T + W_k) C_{k+1}^T + V_{k+1} \big) \nonumber\\[1.75\jot]
	&\qquad \quad \qquad \times \big(Q_{k+1} C_{k+1}^T V_{k+1}^{-1}\big)^T \Big) \nonumber\\[0\jot]
	&\qquad + \tr Q_{k+1}  + \E \Big[V_{k+1}(\breve{e}_{k+1}, R_{k+1}) \big| \breve{e}_k, R_k, u_k \Big],
\end{align}
for $k \in \mathbb{N}_{[0,N-1]}$ with initial condition $V_{N}(\breve{e}_N, R_N) = 0$ together with
\begin{align}
	\breve{e}_{k+1} &= (1 - \bar{\theta}_{k}) A_{k} \breve{e}_{k} + Q_{k+1} C_{k+1}^T V_{k+1}^{-1} \nu_{k+1}, \label{eq:compute:eps}\\[2.5\jot]
	R_{k+1} &= (1 - \bar{\theta}_{k}) A_{k} R_{k} A_{k}^T + \bar{\bar{\theta}}_{k} A_{k} \breve{e}_{k} \breve{e}_{k}^T A_{k}^T. \label{eq:compute:R}
\end{align}
Note that $Q_k$ in (\ref{eq:Vku}), (\ref{eq:compute:eps}), and (\ref{eq:compute:R}) for $k \in [0,N]$ can be computed deterministically forward in time according to
\begin{align}\label{eq:compute:Q}
	Q_{k} &= \big( (A_{k-1} Q_{k-1} A_{k-1}^T \!+\! W_{k-1})^{-1} \!+\! C_k^T V_k^{-1} C_k \big)^{-1},
\end{align}
with initial condition $Q_0 = (M_0^{-1} + C_{0}^T V_{0}^{-1} C_{0})^{-1}$. To evaluate the expected cost-to-go term
\begin{align}
	X_k^{(u_k)} := \E \Big[ V_{k+1}(\breve{e}_{k+1}, R_{k+1}) \big| \breve{e}_k, R_k, u_k \Big],
\end{align}
we can employ a Monte Carlo integration technique based on different realisations of $\nu_{k+1}$, $\theta_k$, and $\gamma_k$. Note that $\nu_{k+1}$ is a Gaussian random variable with zero mean and covariance $C_{k+1} (A_k Q_k A_k^T + W_k) C_{k+1}^T + V_{k+1}$, and $\gamma_k$ and $\theta_k$ are binary random variables with success probabilities $\lambda^c$ and $\rho^c$. To that end, we should generate a number of independent samples of these random variables, and then obtain the Monte Carlo estimate of the expectation by averaging over all samples. 

\begin{algorithm}[t]
\caption{Recursive Computation of $\chi_k(\breve{e}_k, R_k)$}\label{eq:alg}
\begin{algorithmic}[1]
\State \textbf{Input:} Networked system's parameters: time horizon $N$, source matrices $\{A_k, C_k\}$, noise covariances $\{W_k, V_k\}$, initial moments $\{m_0, M_0\}$, error covariance $Q_0$, forward and backward error rates $\{ \lambda, \rho\}$, weighting coefficient $\alpha_k$, for $k \in [0, N]$.
\Statex \textbf{Forward Recursion:}
\For{$k = 1$ up to $N$}
\State Compute $Q_k$ via (\ref{eq:compute:Q}).
\EndFor
\Statex \textbf{Backward Recursion:}
\State \textbf{Initialise:} Define grid points for $\breve{e}_k$ and $R_k$. Set $V_N(\breve{e}_N, R_N) = 0$ for all grid points.
\For{$k = N-1$ down to $0$}
\For{each grid point $(\breve{e}_k, R_k)$}
\For{$u_k \in \{0, 1\}$}
\State Generate Monte Carlo samples of $\nu_{k+1}, \theta_k, \gamma_k$. 
\State Predict $\breve{e}_{k+1}$ for all samples via (\ref{eq:compute:eps}). 
\State Predict $R_{k+1}$ for all samples via (\ref{eq:compute:R}).
\State Compute $X_k^{(u_k)}$ using Monte Carlo integration.
\State Compute $V_k^{(u_k)}$ via (\ref{eq:Vku}).
\EndFor
\State Set $V_k(\breve{e}_k, R_k) = \min\{V_k^{(0)}, V_k^{(1)}\}$.
\State Set $\chi_k(\breve{e}_k, R_k) = \lambda^c \breve{e}_k^T A_k^T A_k \breve{e}_k + \lambda^c \text{tr}(A_k R_k A_k^T) + X_k^{(0)} - X_k^{(1)}$.
\EndFor
\EndFor
\end{algorithmic}\label{eq:alg}
\end{algorithm}

Algorithm~\ref{eq:alg} accordingly presents the key steps in the computation of $\chi_k(\breve{e}_k, R_k)$, an effective term associated with the value function $V_k(\breve{e}_k, R_k)$ used in the policy profile $(\epsilon^\star, \delta^\star)$. The following corollary specifies the complexity of the overall computation.
\begin{corollary}\label{cor1}
Let $\breve{e}_k$ and $R_k$ be discretised in grids with $d_1^{n}$ and $d_2^{n(n+1)/2}$ points, respectively, and $\chi_k(\breve{e}_k, R_k)$ be evaluated based on Monte Carlo integration with $d_0$ samples. The complexity of computing the policy profile $(\epsilon^\star, \delta^\star)$ in Theorem~\ref{thm:1} is $\mathcal{O}(N d_0 d_1^n d_2^{n(n+1)/2})$.
\end{corollary}
\begin{IEEEproof}
We have shown that the value function is a function of $\breve{e}_k$ and $R_k$. Hence, the computational complexity of the policy profile $(\epsilon^\star, \delta^\star)$ is the same as that of solving $(\ref{eq:compute:V})$ for $k \in \mathbb{N}_{[0,N-1]}$. Accordingly, the overall computational complexity is obtained as $\mathcal{O}(N d_0 d_1^n d_2^{n(n+1)/2})$.
\end{IEEEproof}

\begin{remark}
While the recursive formulation explained above facilitates a systematic backward-in-time computation of the value function, the dimensionality of this method may grow rapidly. In practice, efficient storage, pruning of unlikely states, multi-point interpolation, and parallel evaluation of Monte Carlo samples can substantially improve performance. Furthermore, one may use multi-step lookahead methods (see \cite{bertsekas1995DP}), in which instead of the entire future stages only a limited number of future stages are evaluated at each time. For instance, using one-step lookahead policy, we obtain $\chi_k(\breve{e}_k, R_k) \simeq \lambda^c \breve{e}_k^T A_k^T A_k \breve{e}_k + \lambda^c \tr(A_k R_k A_k^T)$. These approximations are particularly valuable for real-time implementations.
\end{remark}

\section{Special Cases}\label{sec:specialcases}
In this section, we delve into several special cases that arise from our main theoretical results. Our goal is to illustrate how the general framework developed in the previous sections applies under different operating regimes. More specifically, we present specialised results for the following cases: reliable channels without feedback, unreliable channels with reliable feedback, unreliable  channels without feedback, and scalar sources. By examining each of these regimes separately, we provide deeper insight into how channel reliability, feedback, and source dimensionality affect the main results.

\subsection{Reliable Channels without Feedback}
This regime models an ideal communication setting where the forward channel delivers all packets successfully, and there are no acknowledgement messages from the receiver. Since the forward channel guarantees delivery, feedback is unnecessary. In this regime, characterised by $\lambda = 0$ and $\rho = 1$, the information sets of the encoder and the decoder are specified~as
\begin{align}
	\mathcal{I}^e_k &= \Big\{ y_t, z^f_t, u_s \big| t \in \mathbb{N}_{[0,k]}, s \in \mathbb{N}_{[0,k-1]} \Big\}, \\[1\jot]
	\mathcal{I}^d_k &= \Big\{ z^f_t, u_s \big| t \in \mathbb{N}_{[0,k]}, s \in \mathbb{N}_{[0,k-1]} \Big\}.
\end{align}

Under this setting, the optimal policy profile and the complexity of its computation are provided by the following proposition and corollary.
\begin{proposition}\label{prop:1}\emph{
The causal tradeoff between the packet rate and the mean square error under reliable communication without acknowledgements admits a globally optimal solution $(\epsilon^\star, \delta^\star)$ such that $\epsilon^\star$ is a symmetric threshold-based scheduling policy given~by
\begin{align}
	u_k = \mathds{1}_{\chi_k - \alpha_k \geq 0},
\end{align}
for $k \in \mathbb{N}_{[0,N-1]}$, in conjunction with $\check{x}_k = \E[x_k | \mathcal{I}^e_k]$, where $\chi_k(\tilde{e}_k) = \tilde{e}_k^T A_k^T A_k \tilde{e}_k + \Delta_k$ is a symmetric function of $\tilde{e}_k$, which requires solving
\begin{align}
	\check{x}_{k} &= A_{k-1} \check{x}_{k-1} + K_{k} \nu_k, \\[1.75\jot]
	Q_{k} &= \big( (A_{k-1} Q_{k-1} A_{k-1}^T + W_{k-1})^{-1} + L_k \big)^{-1}, \\[2.5\jot]
	\tilde{e}_k &= (1- u_{k-1} ) A_{k-1} \tilde{e}_{k-1} + K_k \nu_k, \label{eq:type1:et}	
\end{align}
for $k \in \mathbb{N}_{[1,N]}$ with initial conditions $\check{x}_0 = m_0 + K_0 \nu_0$, $Q_0 = (M_0^{-1} + C_{0}^T V_{0}^{-1} C_{0})^{-1}$, and $\tilde{e}_0 = K_0 \nu_0$, where $\nu_k = y_k - C_k A_{k-1} \check{x}_{k-1}$, $K_{k} = Q_{k} C_{k}^T V_{k}^{-1}$, and $L_k = C_{k}^T V_{k}^{-1} C_{k}$; and $\delta^\star$ is a linear switching estimation policy given~by
\begin{align}
	\hat{x}_k = A_{k-1} \hat{x}_{k-1} + u_{k-1} A_{k-1} \tilde{e}_{k-1}, \label{eq:type1:xh}
\end{align}
for $k \in \mathbb{N}_{[1,N]}$ with initial condition $\hat{x}_0 = m_0$, where $\hat{x}_k = \E[x_k | \mathcal{I}^d_k]$, without being influenced by implicit information.}
\end{proposition}
\begin{IEEEproof}
These results follow directly from Theorem~1 by substituting $\lambda^c = 1$, $\gamma_k = 1$, and $\theta_k = 0$ for $k \in \mathbb{N}_{[0,N]}$. Following these transformations, $\breve{e}_k$ reduces to $\tilde{e}_k$ and $R_k$ becomes zero for $k \in \mathbb{N}_{[0,N]}$.
\end{IEEEproof}

\begin{corollary}\label{cor2}
Let $\tilde{e}_k$ be discretised in a grid with $d_1^{n}$ points, and $\chi_k$ be evaluated based on Monte Carlo integration with $d_0$ samples. The complexity of computing the policy profile $(\epsilon^\star, \delta^\star)$ in Proposition~\ref{prop:1} is $\mathcal{O}(N d_0 d_1^n)$.
\end{corollary}
\begin{IEEEproof}
In this regime, the value function depends only on $\tilde{e}_k$. Hence, the computational complexity of the policy profile $(\epsilon^\star, \delta^\star)$ is the same as that of solving $(\ref{eq:compute:V})$ given $\breve{e}_k = \tilde{e}_k$ and $R_k = 0$, for $k \in \mathbb{N}_{[0,N-1]}$. Accordingly, the overall computational complexity is obtained as $\mathcal{O}(N d_0 d_1^n)$.
\end{IEEEproof}

\begin{remark}
The best possible system performance among all considered communication regimes is achieved in this ideal regime. The scheduling policy retains a symmetric threshold-based structure and depends solely on the estimation mismatch at each time, while the estimation policy updates state estimates more frequently than in the general regime because all transmitted packets are received. The evolution of the estimation mismatch is corrected only through scheduling decisions, and the absence of acknowledgements has no adverse effect due to the perfect reliability of the forward channel. The computational complexity is relatively low, reflecting the efficiency of optimal policy computation in~this~regime.
\end{remark}

\subsection{Unreliable Channel with Reliable Feedback}
This regime models a communication setting in which the forward channel is subject to losses, while the receiver can reliably transmit acknowledgement messages over a zero-delay backward channel. Such a configuration is representative of many practical systems, including wireless control systems with robust and dependable feedback links. In this regime, characterised by $\lambda > 0$ and $\rho = 0$, the information sets of the encoder and the decoder are specified~as
\begin{align}
	\mathcal{I}^e_k &= \Big\{ y_t, z^f_t, z^b_t, u_s, \gamma_s \big| t \in \mathbb{N}_{[0,k]}, s \in \mathbb{N}_{[0,k-1]} \Big\}, \\[1\jot]
	\mathcal{I}^d_k &= \Big\{ z^f_t, z^b_t, u_s, \gamma_s \big| t \in \mathbb{N}_{[0,k]}, s \in \mathbb{N}_{[0,k-1]} \Big\}.
\end{align}

Under this setting, the optimal policy profile and the complexity of its computation are provided by the following proposition and corollary.
\begin{proposition}\label{prop:2}\emph{
The causal tradeoff between the packet rate and the mean square error under unreliable communication with reliable acknowledgements admits a globally optimal solution $(\epsilon^\star, \delta^\star)$ such that $\epsilon^\star$ is a symmetric threshold-based scheduling policy given~by
\begin{align}
	u_k = \mathds{1}_{\chi_k - \alpha_k \geq 0},
\end{align}
for $k \in \mathbb{N}_{[0,N-1]}$, in conjunction with $\check{x}_k = \E[x_k | \mathcal{I}^e_k]$, where $\chi_k(\tilde{e}_k) = \lambda^c \tilde{e}_k^T A_k^T A_k \tilde{e}_k + \Delta_k$ is a symmetric function of $\tilde{e}_k$, which requires solving
\begin{align}
	\check{x}_{k} &= A_{k-1} \check{x}_{k-1} + K_{k} \nu_k, \\[1.75\jot]
	Q_{k} &= \big( (A_{k-1} Q_{k-1} A_{k-1}^T + W_{k-1})^{-1} + L_k \big)^{-1}, \\[2.5\jot]
	\tilde{e}_k &= (1- u_{k-1} \gamma_{k-1}) A_{k-1} \tilde{e}_{k-1} + K_k \nu_k, \label{eq:type2:et}
\end{align}
for $k \in \mathbb{N}_{[1,N]}$ with initial conditions $\check{x}_0 = m_0 + K_0 \nu_0$, $Q_0 = (M_0^{-1} + C_{0}^T V_{0}^{-1} C_{0})^{-1}$, and $\tilde{e}_0 = K_0 \nu_0$, where $\nu_k = y_k - C_k A_{k-1} \check{x}_{k-1}$, $K_{k} = Q_{k} C_{k}^T V_{k}^{-1}$, and $L_k = C_{k}^T V_{k}^{-1} C_{k}$; and $\delta^\star$ is a linear switching estimation policy given by
\begin{align}
	\hat{x}_k = A_{k-1} \hat{x}_{k-1} + u_{k-1} \gamma_{k-1} A_{k-1} \tilde{e}_{k-1}, \label{eq:type2:xh}
\end{align}
for $k \in \mathbb{N}_{[1,N]}$ with initial condition $\hat{x}_0 = m_0$, where $\hat{x}_k = \E[x_k | \mathcal{I}^d_k]$, without being influenced by implicit information.}
\end{proposition}
\begin{IEEEproof}
These results follow directly from Theorem~1 by substituting $\theta_k = 1$ for $k \in \mathbb{N}_{[0,N]}$. Following this transformation, $\breve{e}_k$ reduces to $\tilde{e}_k$ and $R_k$ becomes zero for $k \in \mathbb{N}_{[0,N]}$.
\end{IEEEproof}

\begin{corollary}\label{cor3}
Let $\tilde{e}_k$ be discretised in a grid with $d_1^{n}$ points, and $\chi_k$ be evaluated based on Monte Carlo integration with $d_0$ samples. The complexity of computing the policy profile $(\epsilon^\star, \delta^\star)$ in Proposition~\ref{prop:2} is $\mathcal{O}(N d_0 d_1^n)$.
\end{corollary}
\begin{IEEEproof}
In this regime, the value function depends only on $\tilde{e}_k$. Hence, the computational complexity of the policy profile $(\epsilon^\star, \delta^\star)$ is the same as that of solving $(\ref{eq:compute:V})$ given $\breve{e}_k = \tilde{e}_k$ and $R_k = 0$ for $k \in \mathbb{N}_{[0,N-1]}$. Accordingly, the overall computational complexity is obtained as $\mathcal{O}(N d_0 d_1^n)$.
\end{IEEEproof}

\begin{remark}
In this regime, the presence of reliable feedback enables the encoder to update its local belief more accurately than in the general regime. The scheduling policy retains a symmetric threshold-based structure and incorporates delivery outcomes, while the estimation policy follows the same structure as in the general regime. The feedback introduces a form of compensation that improves scheduling decisions. Although communication losses degrade performance compared to the ideal regime, reliable acknowledgements allow the system to recover some of the lost performance relative to the general regime. The computational complexity remains manageable and identical to the ideal regime, owing to the feedback aiding in estimation mismatch correction.
\end{remark}

\subsection{Unreliable Channels without Feedback}
This regime models the most constrained communication setting among all considered communication regimes in which the forward channel is subject to losses and there are no acknowledgement messages from the receiver. Such a configuration might arise in lightweight Internet-of-Things devices where feedback is unavailable. In this regime, characterised by $\lambda > 0$ and $\rho = 1$, the information sets of the encoder and the decoder are specified~as 
\begin{align}
	\mathcal{I}^e_k &= \Big\{ y_t, u_s \big| t \in \mathbb{N}_{[0,k]}, s \in \mathbb{N}_{[0,k-1]} \Big\}, \\[1\jot]
	\mathcal{I}^d_k &= \Big\{ z^f_t, u_s, \gamma_s \big| t \in \mathbb{N}_{[0,k]}, s \in \mathbb{N}_{[0,k-1]} \Big\}.
\end{align}

Under this setting, the optimal policy profile and the complexity of its computation are provided by the following proposition and corollary.
\begin{proposition}\label{prop:3}\emph{
The causal tradeoff between the packet rate and the mean square error under unreliable communication without acknowledgements admits a globally optimal solution $(\epsilon^\star, \delta^\star)$ such that $\epsilon^\star$ is a symmetric threshold-based scheduling policy given~by
\begin{align}
	u_k = \mathds{1}_{\chi_k - \alpha_k \geq 0},
\end{align}
for $k \in \mathbb{N}_{[0,N-1]}$, in conjunction with $\check{x}_k = \E[x_k | \mathcal{I}^e_k]$, where $\chi_k(\breve{e}_k, R_k) = \lambda^c \breve{e}_k^T A_k^T A_k \breve{e}_k + \lambda^c \tr(A_k R_k A_k^T) + \Delta_k$ is a symmetric function of $\breve{e}_k$ and also a function of $R_k$, which requires solving
\begin{align}
	\check{x}_{k} &= A_{k-1} \check{x}_{k-1} + K_{k} \nu_k,\\[1.75\jot]
	Q_{k} &= \big( (A_{k-1} Q_{k-1} A_{k-1}^T + W_{k-1})^{-1} + L_k \big)^{-1}, \\[2\jot]
	\breve{e}_k &= (1 - \bar{\theta}_{k-1}) A_{k-1} \breve{e}_{k-1} + K_k \nu_k, \label{eq:type3:eps}\\[2\jot]
	R_k &= (1 - \bar{\theta}_{k-1}) A_{k-1} R_{k-1} A_{k-1}^T  + \bar{\bar{\theta}}_{k-1} H_{k-1}, \label{eq:type3:R}	
\end{align}
for $k \in \mathbb{N}_{[1,N]}$ with initial conditions $\check{x}_0 = m_0 + K_0 \nu_0$, $Q_0 = (M_0^{-1} + C_{0}^T V_{0}^{-1} C_{0})^{-1}$, $\breve{e}_0 = K_0 \nu_0$, and $R_0 = 0$, where $\nu_k = y_k - C_k A_{k-1} \check{x}_{k-1}$, $K_{k} = Q_{k} C_{k}^T V_{k}^{-1}$, $L_k = C_{k}^T V_{k}^{-1} C_{k}$, $\bar{\theta}_{k-1} = \lambda^c u_{k-1}$, $\bar{\bar{\theta}}_{k-1} = (\lambda^c - {\lambda^c}^2) u_{k-1}$, and $H_{k-1} = A_{k-1} \breve{e}_{k-1} \breve{e}_{k-1}^T A_{k-1}^T$; and $\delta^\star$ is a linear switching estimation policy given by
\begin{align}
	\hat{x}_k = A_{k-1} \hat{x}_{k-1} + u_{k-1} \gamma_{k-1} A_{k-1} \tilde{e}_{k-1}, \label{eq:type3:xh}
\end{align}
for $k \in \mathbb{N}_{[1,N]}$ with initial condition $\hat{x}_0 = m_0$, where $\hat{x}_k = \E[x_k | \mathcal{I}^d_k]$, without being influenced by implicit information.}
\end{proposition}
\begin{IEEEproof}
These results follow directly from Theorem~1 by substituting $\theta_k = 0$ for $k \in \mathbb{N}_{[0,N]}$. Following this transformation, we should update $\bar{\theta}_{k-1} = \lambda^c u_{k-1}$ and $\bar{\bar{\theta}}_{k-1} = \lambda^c u_{k-1} - {\lambda^c}^2 u_{k-1}$.
\end{IEEEproof}

\begin{corollary}\label{cor4}
Let $\breve{e}_k$ and $R_k$ be discretised in grids with $d_1^{n}$ and $d_2^{n(n+1)/2}$ points, respectively, and $\chi_k$ be evaluated based on Monte Carlo integration with $d_0$ samples. The complexity of computing the policy profile $(\epsilon^\star, \delta^\star)$ in Proposition~\ref{prop:3} is $\mathcal{O}(N d_0 d_1^n d_2^{n(n+1)/2})$.
\end{corollary}
\begin{IEEEproof}
As in the general regime, the value function is a function of $\breve{e}_k$ and $R_k$. Hence, the computational complexity of the policy profile $(\epsilon^\star, \delta^\star)$ is the same as that of solving $(\ref{eq:compute:V})$ for $k \in \mathbb{N}_{[0,N-1]}$. Accordingly, the overall computational complexity is obtained as $\mathcal{O}(N d_0 d_1^n d_2^{n(n+1)/2})$.
\end{IEEEproof}

\begin{remark}
This regime is expected to yield the worst system performance among all considered communication regimes. The scheduling policy preserves a symmetric threshold-based structure and decides whether to transmit a packet at each time without knowing the success of prior transmissions, while the estimation policy follows the same structure as in the general regime. The complete absence of feedback introduces significant challenges here, as the encoder is now forced to operate blindly. This means that the encoder must estimate the evolution of the estimation mismatch with uncertainty accumulating over time without any corrections and with a risk of divergence. The computational complexity is comparable to that of the general regime. However, more grid points may be required as uncertainty grows.
\end{remark}

\subsection{Scalar Sources}
This regime specialises the general framework to scalar systems, where the state and the output of the source are one-dimensional. Such a configuration is common in applications like temperature and voltage monitoring in cyber-physical systems. In this regime, characterised by $n = 1$ and $m = 1$, the information sets of the encoder and the decoder can be represented similar to those in (\ref{eq:infoset1}) and (\ref{eq:infoset2}).

\begin{figure}[t]
\center
  \includegraphics[width= 0.97\linewidth]{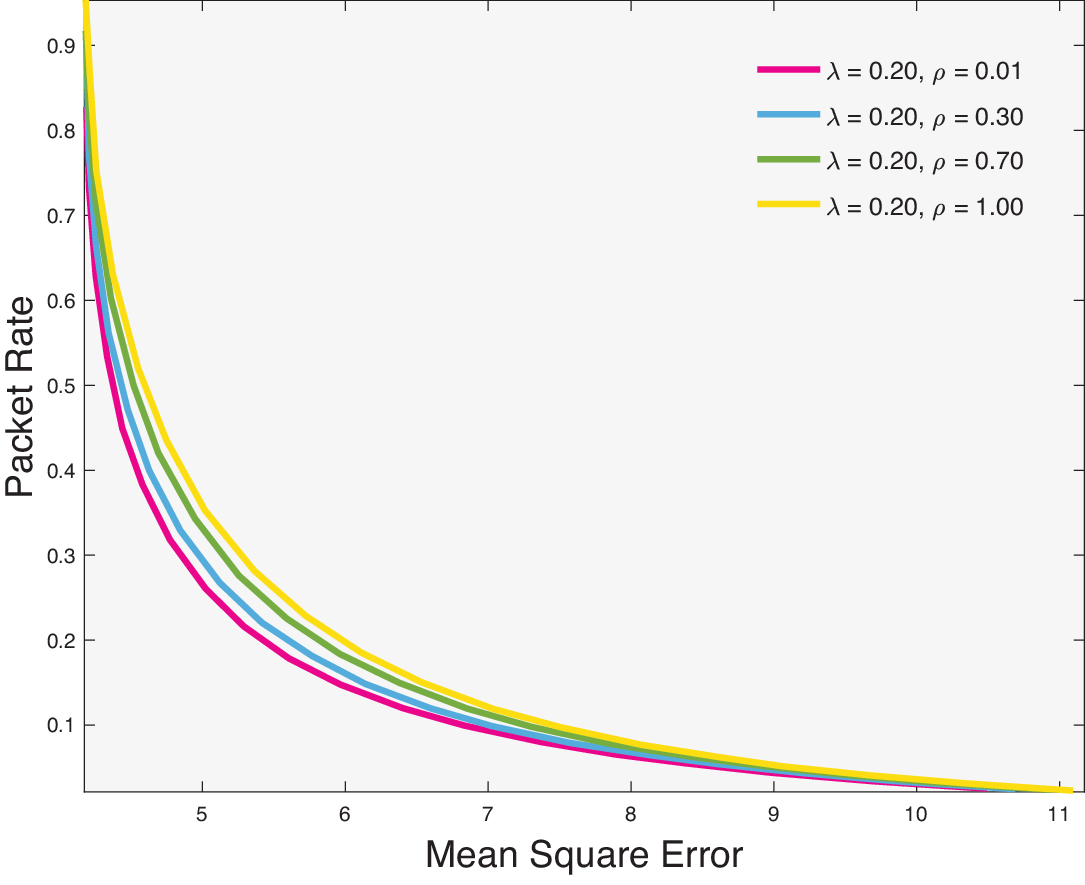}
  \caption{Causal tradeoff curves for remote state estimation when $\lambda = 0.20$ and $\rho \in \{0.01, 0.30, 0.70, 1.00\}$.}
  \label{fig:tradeoff1}
\end{figure}

\begin{figure}[t]	
\center
  \includegraphics[width= 0.97\linewidth]{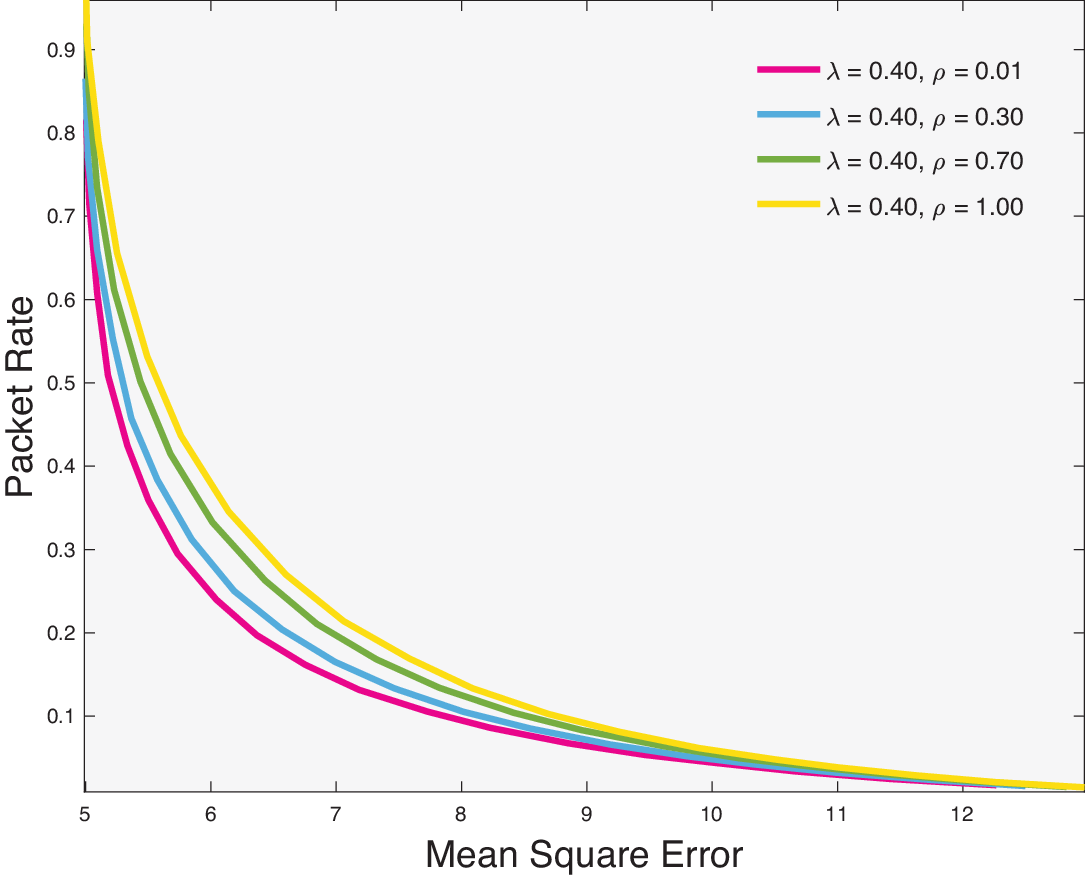}
  \caption{Causal tradeoff curves for remote state estimation when $\lambda = 0.40$ and $\rho \in \{0.01, 0.30, 0.70, 1.00\}$.}
  \label{fig:tradeoff2}
\end{figure}

\begin{figure}[t]	
\center
  \includegraphics[width= 0.97\linewidth]{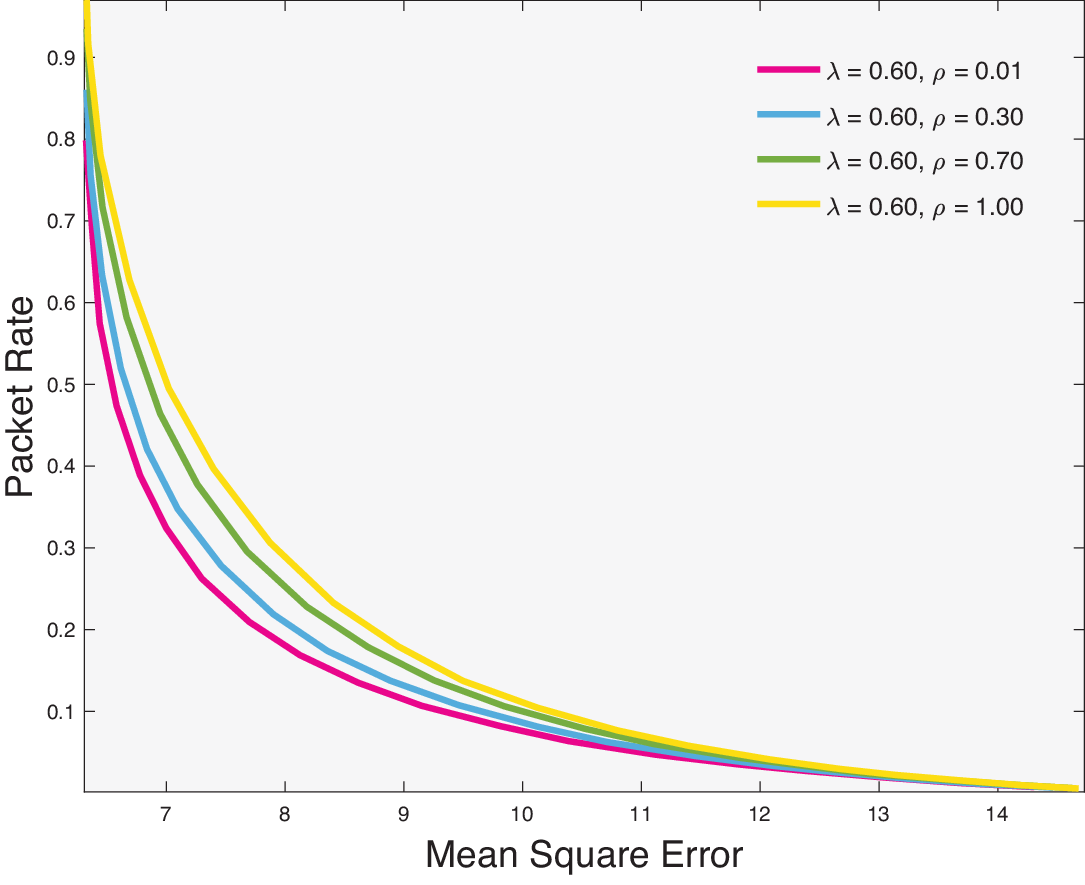}
  \caption{Causal tradeoff curves for remote state estimation when $\lambda = 0.60$ and $\rho \in \{0.01, 0.30, 0.70, 1.00\}$.}
  \label{fig:tradeoff3}
\end{figure}

\begin{figure}[t]
\vspace{2.4mm}
\center
  \includegraphics[width= 0.97\linewidth]{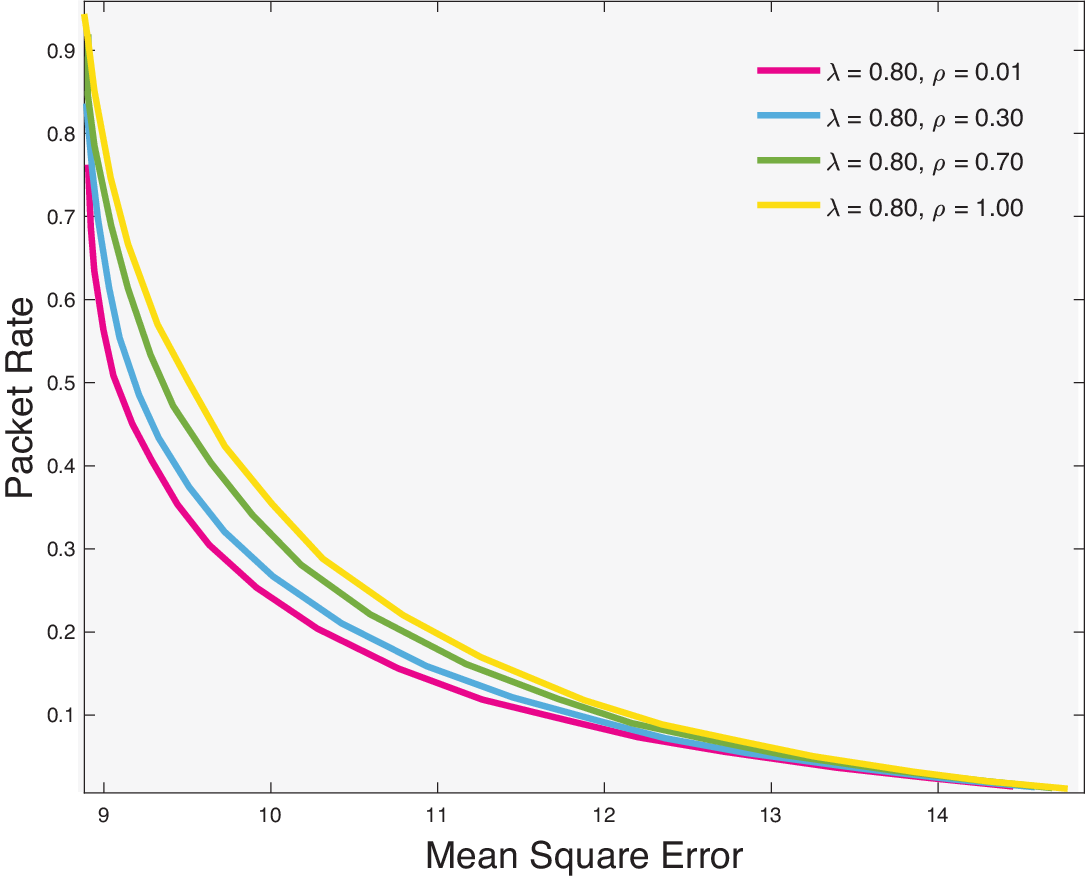}
  \caption{Causal tradeoff curves for remote state estimation when $\lambda = 0.80$ and $\rho \in \{0.01, 0.30, 0.70, 1.00\}$.}
  \label{fig:tradeoff4}
\end{figure}

\begin{figure*}[t!]
\centering
  \includegraphics[width=.999\linewidth]{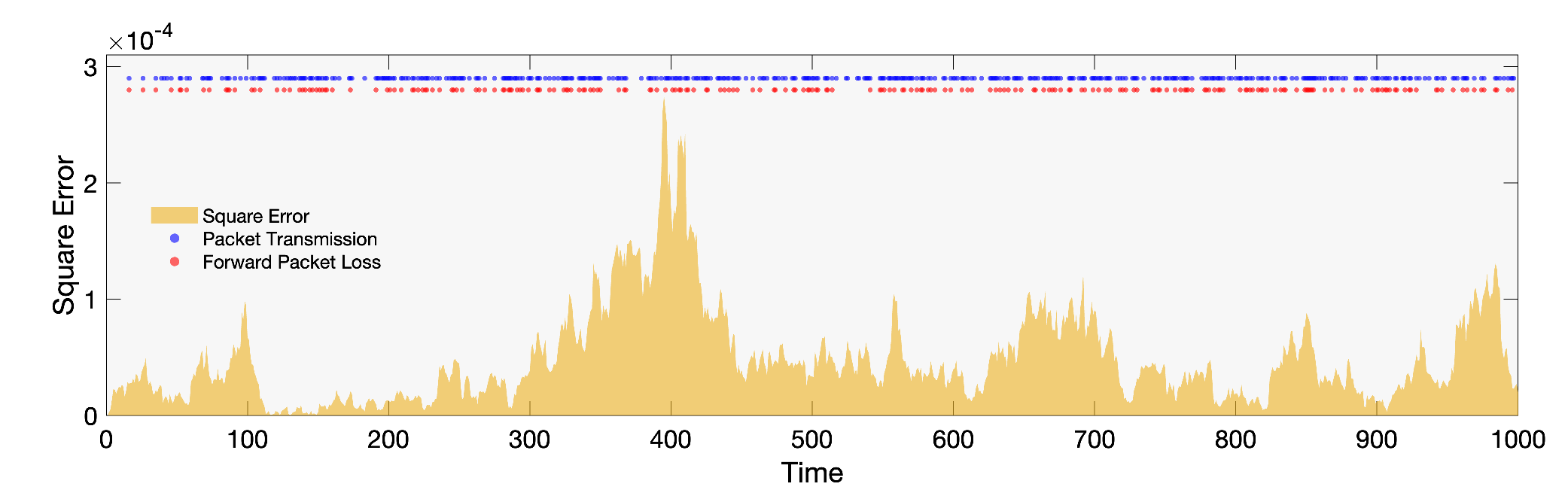}
  \caption{Key trajectories in remote state estimation over a highly unreliable channel without feedback.}
  \label{fig:sat1}
\end{figure*}

\begin{figure*}[t!]
\centering
  \includegraphics[width=.999\linewidth]{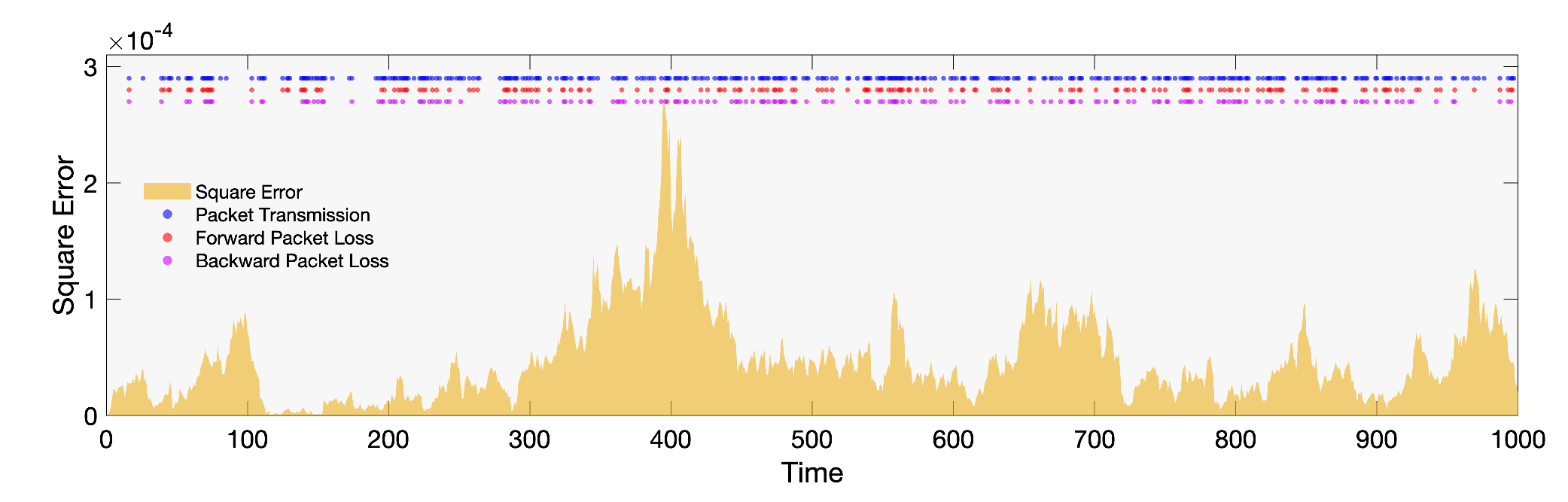}
  \caption{Key trajectories in remote state estimation over a highly unreliable channel with unreliable feedback.}
  \label{fig:sat2}
\end{figure*}

Under this setting, the optimal policy profile and the complexity of its computation are provided by the following proposition and corollary.
\begin{proposition}\label{prop:4}\emph{
The causal tradeoff between the packet rate and the mean square error for a scalar source under unreliable communication with unreliable acknowledgements admits a globally optimal solution $(\epsilon^\star, \delta^\star)$ such that $\epsilon^\star$ is a symmetric threshold-based scheduling policy given~by
\begin{align}
	u_k = \mathds{1}_{\chi_k - \alpha_k \geq 0},
\end{align}
for $k \in \mathbb{N}_{[0,N-1]}$, in conjunction with $\check{x}_k = \E[x_k | \mathcal{I}^e_k]$, where $\chi_k(\breve{e}_k, R_k) = \lambda^c A_k^2 \breve{e}_k^2 + \lambda^c A_k^2 R_k + \Delta_k$ is a symmetric function of $\breve{e}_k$ and also a function of $R_k$, which requires solving
\begin{align}
	\check{x}_{k} &= A_{k-1} \check{x}_{k-1} + \frac{Q_{k} C_{k}}{V_{k}} \nu_k, \label{eq:type4:xc}\\[1.5\jot]
	Q_{k} &= \frac{1}{\frac{1}{A_{k-1}^2 Q_{k-1} + W_{k-1}} + \frac{C_{k}^2}{V_{k}}} \: ,\\[1\jot]
	\breve{e}_k &= (1 - \bar{\theta}_{k-1}) A_{k-1} \breve{e}_{k-1} + \frac{Q_{k} C_{k}}{V_{k}} \nu_k,\label{eq:type4:eps}\\[2\jot]
	R_k &= (1 - \bar{\theta}_{k-1}) A_{k-1}^2 R_{k-1} + \bar{\bar{\theta}}_{k-1} A_{k-1}^2 \breve{e}_{k-1}^2, \label{eq:type4:R}
\end{align}
for $k \in \mathbb{N}_{[1,N]}$ with initial conditions $\check{x}_0 = m_0 + K_0 \nu_0$, $Q_0 = 1/(1/M_0 + C_{0}^2/V_{0})$, $\breve{e}_0 = Q_{0} C_{0} \nu_0 /V_{0}$, and $R_0 = 0$, where $\nu_k = y_k - C_k A_{k-1} \check{x}_{k-1}$, $\bar{\theta}_{k-1} = \theta_{k-1} u_{k-1} \gamma_{k-1} + \lambda^c (1 - \theta_{k-1}) u_{k-1}$, and $\bar{\bar{\theta}}_{k-1} = (\lambda^c - {\lambda^c}^2) (1 - \theta_{k-1}) u_{k-1}$; and $\delta^\star$ is a linear switching estimation policy given by
\begin{align}
	\hat{x}_k = A_{k-1} \hat{x}_{k-1} + u_{k-1} \gamma_{k-1} A_{k-1} \tilde{e}_{k-1}, \label{eq:type4:xh}
\end{align}
for $k \in \mathbb{N}_{[1,N]}$ with initial condition $\hat{x}_0 = m_0$, where $\hat{x}_k = \E[x_k | \mathcal{I}^d_k]$, without being influenced by implicit information.}
\end{proposition}
\begin{IEEEproof}
These results follow directly from Theorem~1 by writing all equations in a scalar form.
\end{IEEEproof}

\begin{corollary}\label{cor5}
Let $\breve{e}_k$ and $R_k$ be discretised in grids with $d_1$ and $d_2$ points, respectively, and $\chi_k$ be evaluated based on Monte Carlo integration with $d_0$ samples. The complexity of computing the policy profile $(\epsilon^\star, \delta^\star)$ in Proposition~\ref{prop:4} is $\mathcal{O}(N d_0 d_1 d_2)$.
\end{corollary}
\begin{IEEEproof}
As in the general regime, the value function remains a function of $\breve{e}_k$ and $R_k$. Hence, the computational complexity of the policy profile $(\epsilon^\star, \delta^\star)$ is the same as that of solving the scalar version of $(\ref{eq:compute:V})$ for $k \in \mathbb{N}_{[0,N-1]}$. Accordingly, the overall computational complexity is obtained as $\mathcal{O}(N d_0 d_1 d_2)$.
\end{IEEEproof}

\begin{remark}
In the scalar regime, matrix-valued quantities reduce to scalars, and inner products become simple multiplications. This leads to a substantial simplification of the recursive equations and the computation of intermediate terms. Moreover, the computational complexity of the optimal policy profile decreases significantly, as the discretisation now only spans one-dimensional spaces. Consequently, scalar systems can serve as a testbed for exploring advanced design strategies before extending them to higher-dimensional settings. In spite of their simplicity, scalar systems carry the key elements of the general theory, making them valuable for theoretical validation and practical prototyping.
\end{remark}

\section{Numerical Analysis}\label{sec:example}
In this section, we present two illustrative examples that demonstrate the applicability of the theoretical results developed in the paper. The first example involves an industrial machine whose temperature should be monitored at a data centre, while the second example involves a spin-stabilised spacecraft whose angular velocity should be tracked at a ground station. These examples are chosen to reflect practical remote estimation scenarios where communication constraints play a critical~role.

\subsection{Remotely Tracking Temperature of a Machine}
In this example, we consider an industrial machine whose temperature deviation from a known reference temperature evolves according to
\begin{align}\label{eq:example1}
	\frac{dT}{dt} = -c T + \zeta,
\end{align}
where $T$ is the temperature deviation, $c$ is the heat coefficient, and $\zeta$ is a Gaussian noise. We assume that $c = 1.054$. A sensor deployed in the machine measures its temperature. The sensory information should be transmitted over a costly communication channel to a data centre where the temperature is estimated in real-time. The dynamic equation in (\ref{eq:example1}) can be discretised over the time horizon of the experiment $N = 1000$. The discrete-time state and output equations of the form (\ref{eq:sys}) and (\ref{eq:sens}) are specified by $A_k = 0.9$, $C_k = 1$, $W_k = 3$, $V_k = 1$, for $k \in \mathbb{N}_{[0,N]}$, $m_0 = 0$, and $M_0 = 1$. The forward error rate $\lambda$, the backward error rate $\rho$, and the weighting coefficient $\alpha_k$ are assumed to be constant over time. We used Monte Carlo simulation, and analysed how feedback influences the performance limits of the networked system in various communication regimes. Figs.~\ref{fig:tradeoff1}–\ref{fig:tradeoff4} depict the causal tradeoff curves when $\lambda$ is equal to $0.20$, $0.40$, $0.60$, and $0.80$, respectively, while $\rho$ takes values in the set $\{0.01, 0.30, 0.70, 1.00\}$. These tradeoff curves delineate the boundaries of achievable regions: any point above a given curve corresponds to a feasible combination of communication cost and estimation error, whereas points below the curve are unattainable under the given condition. From these plots, several key insights emerge. First, for each fixed value of $\lambda$, decreasing $\rho$ (i.e., improving feedback quality) leads to a larger achievable region, enabling better estimation performance with similar communication effort. Second, when $\rho = 0.01$ (i.e., feedback is nearly perfect), the encoder has an accurate knowledge of the decoder's belief, which enhances scheduling decisions and leads to the best system performance. Third, as $\rho$ increases toward $1$ (i.e., feedback becomes unavailable), the encoder receives little useful feedback, resulting in degraded performance and greater packet rates. Lastly, as the forward error rate $\lambda$ increases, the estimation performance degrades and curves shift to the right for all values of $\rho$, illustrating the critical role of reliable forward communication in maintaining estimation quality. These observations underscore the importance of robust communication mechanisms in remote estimation systems and highlight how the optimal policies can adaptively balance communication and estimation objectives even under significant uncertainty.

\subsection{Remotely Tracking Angular Velocity of a Spacecraft}
In this example, we consider a spin-stabilised spacecraft whose body is spinning about the axis of symmetry, with a constant angular velocity $\omega_z = \omega_0$. For such a spacecraft, the Euler equation is written as
\begin{align}\label{eq:example2}\setlength\arraycolsep{6pt}\def\arraystretch{1.5}
\begin{bmatrix}
 \frac{d \omega_x}{dt}\\
 \frac{d \omega_y}{dt}\\
 \frac{d \omega_z}{dt}    
 \end{bmatrix}
 =
\begin{bmatrix}
 0 & \frac{I_y - I_z}{I_x} \omega_0 & 0\\
 \frac{I_z - I_x}{I_y} \omega_0 & 0 & 0 \\
 0 & 0 & 0   
 \end{bmatrix}
\begin{bmatrix}
 \omega_x\\
 \omega_y\\
 \omega_z     
 \end{bmatrix} 
+
\begin{bmatrix}
e_x\\
e_y \\
e_z	
\end{bmatrix},
\end{align}
where $(\omega_x,\omega_y,\omega_z)$ is the angular velocity, $(I_x,I_y,I_z)$ is the moment of inertia, and $(e_x,e_y,e_z)$ is a Gaussian disturbance torque acting on the spacecraft. Note that for spin stability, the spacecraft must be spinning either about the major or minor axis of inertia. We assume that $I_x = I_y = 20 \ \text{kg.m$^2$}$, $I_z = 200 \ \text{kg.m$^2$}$, and $\omega_0 = 4 \pi \ \text{rad/s}$. A sensor in the spacecraft measures each component of the angular velocity. The sensory information should be transmitted over a costly communication channel to a ground station where the angular velocity is estimated in real-time. The Euler equation in (\ref{eq:example2}) can be discretised over the time horizon of the experiment $N = 1000$. The discrete-time state and output equations of the form (\ref{eq:sys}) and (\ref{eq:sens}) are specified by $A_k = [0.4258, -0.9048, 0; 0.9048, 0.4258, 0; 0, 0 , 1]$, $W_k = 10^{-6} \diag\{0.2245,0.2245,0.0025\}$, $C_k = \diag \{1,1,1\}$, $V_k = 10^{-3} \diag \{ 1, 1, 1\}$, for all $k \in \mathbb{N}_{[0,N]}$, $m_0 = [0;0;2\pi]$, and $M_0 = 10W$. We assume that the weighting coefficient is $\alpha_k = 6 \times 10^{-7}$ for $k \in \mathbb{N}_{[0,N]}$. We studied two communication regimes, and analysed key trajectories of the networked system. The first regime is specified by $\lambda = 0.4$ and $\rho = 1.0$ (i.e., a highly unreliable channel without feedback), while the second regime by $\lambda = 0.4$ and $\rho = 0.4$ (i.e., a highly unreliable channel with unreliable feedback). For a simulated realisation of the system, Fig.~\ref{fig:sat1} shows the mean square error, packet transmission, and forward packet loss trajectories under the first regime. In this case, the total mean square error is $0.0486$, with $443$ total transmissions and $231$ forward packet losses. For the same realisation of the system, Fig.~\ref{fig:sat2} shows the mean square error, packet transmission, forward packet loss, and backward packet loss trajectories under the second regime. In this case, the total mean square error is slightly lower at $0.0479$, with $377$ total transmissions, $188$ forward packet losses, and $189$ backward packet losses. As expected, the lack of a feedback channel impairs the encoder's knowledge of the decoder's belief, leading to more frequent packet transmissions. However, when feedback is available, even unreliably, the number of packet transmissions decreases significantly while maintaining comparable estimation performance. This demonstrates the critical role of feedback in reducing communication cost in remote estimation. For spacecraft operating in deep-space missions, where communication resources are limited and packet loss is common, these findings suggest that incorporating even partial feedback can be substantially beneficial.

\section{Conclusions}\label{sec:conclusion}
In this article, we developed a comprehensive theoretical framework for remote estimation over unreliable channels with unreliable feedback. We formulated a causal tradeoff between the packet rate and the mean square error, and characterised a globally optimal coding policy profile. We proved that the value function can be transformed into a lower-dimensional form. This dimensionality reduction not only enhances interpretability but also leads to tractable implementation. We proposed an algorithm for the computation of the optimal policies, along with an analysis of its computational complexity. To illustrate the scope of the theoretical results, we examined several special cases that emerge as direct consequences of the general framework. In particular, we studied: reliable channels without feedback, unreliable channels with reliable feedback, unreliable channels without feedback, and scalar sources. We also presented a numerical analysis, demonstrating how the proposed policies perform under different regimes. Our numerical results highlighted the impact of channel reliability on remote estimation, underscoring the critical role of feedback in achieving efficient system operation.

We acknowledge that grid-based discretisation inevitably suffers from the curse of dimensionality as the system dimension increases. For high-dimensional systems, exact implementation may therefore become computationally prohibitive. An important direction for future work is the development of scalable approximation methods, such as function-approximation-based value iteration, adaptive gridding, or learning-based approaches, to enable efficient policy synthesis in large-scale settings while preserving the structural insights established in this paper. We further acknowledge that ignoring quantisation error effectively separates rate-distortion considerations from the estimation problem. Relaxing this assumption would necessitate a joint treatment of quantisation, scheduling, and estimation, which may result in significantly greater analytical and computational complexity. Exploring such integrated designs, along with developing tractable approximation methods for quantised communication, constitutes a natural avenue for future research. Several other promising directions remain open. For instance, analysing the impact of transmission delays in both forward and backward channels could provide a more realistic extension of the current framework. Another important topic is the study of multi-user systems where multiple encoders share a common unreliable channel with unreliable feedback, introducing new challenges in scheduling, resource allocation, and coordination among users.

\section*{Appendix: Derivation of Main Results}
This section is dedicated to the derivation of the main results. We begin with the following Lemma, showing that the conditional means $\E[x_k | \mathcal{I}^e_k]$ and $\E[x_k | \mathcal{I}^d_k]$ are instrumental in the operations performed by the encoder and the decoder.

\begin{lemma}\label{lemma0}
Without loss of optimality, at each time $k$, one can adopt $\check{x}_k = \E[x_k | \mathcal{I}^e_k]$ as the message that can be transmitted by the encoder, and $\hat{x}_k = \E[x_k | \mathcal{I}^d_k]$ as the state estimate that can be computed by the decoder.	
\end{lemma}
\begin{IEEEproof}
Let $\epsilon^o$ be a globally optimal scheduling policy that is implemented. Given all information available at the decoder, $\E[x_k | \mathcal{I}^d_k]$ is the optimal value that minimises the mean square error at the decoder at time $k$. Moreover, $\E[x_k | \mathcal{I}^e_k]$ fuses all the previous and current outputs of the process that are available at the encoder at time $k$. This implies that, from the minimum mean-square-error perspective, the decoder is able to develop a state estimate upon the receipt of a message at time $k+1$, that would be the same if it had all the previous outputs of the process until time $k$, which is the best possible case for the decoder. Hence, both claims hold.
\end{IEEEproof}

The next two lemmas characterise the optimal estimators at the encoder and the decoder.
 
\begin{lemma}\label{lem:estimator-encoder}\emph{
The optimal estimator minimising the mean square error at the encoder satisfies the recursive equations
\begin{align}
	\check{x}_{k} &= m_{k} + K_{k} ( y_{k} - C_{k} m_{k}), \label{eq:est-KF-xc} \\[2.25\jot]
	m_{k} &= A_{k-1} \check{x}_{k-1} \label{eq:est-KF-m},
\end{align}
\begin{align}
	Q_{k} &= ( M_{k}^{-1} + C_{k}^T V_{k}^{-1} C_{k} )^{-1}, \\[2.25\jot]
	M_{k} &= A_{k-1} Q_{k-1} A_{k-1}^T + W_{k-1},
\end{align}
for $k \in \mathbb{N}_{[1,N]}$ with initial conditions $\check{x}_0 = m_0 + K_0(y_0 - C_0 m_0)$ and $Q_0 = (M_0^{-1} + C_{0}^T V_{0}^{-1} C_{0})^{-1}$, where $m_k = \E[ x_k | \mathcal{I}^e_{k-1}]$, $K_{k} = Q_{k} C_{k}^T V_{k}^{-1}$, $Q_k = \Cov[x_k | \mathcal{I}^e_{k}]$, and $M_k = \Cov[x_k | \mathcal{I}^e_{k-1}]$.}
\end{lemma}

\begin{IEEEproof}
Observe that, given $\mathcal{I}^e_k$, the minimum mean-square-error estimator is $\E[ x_k | \mathcal{I}^e_k]$. It is evident that this estimator satisfies the Kalman filter equations (see, e.g., \cite{stengel1994}). 
\end{IEEEproof}

\begin{lemma}\label{lem:estimator-decoder}\emph{
The optimal estimator minimising the mean square error at the decoder satisfies the recursive equation
\begin{align}\label{eq:est-monitor}
	\hat{x}_{k} &=  A_{k-1} \hat{x}_{k-1} + u_{k-1} \gamma_{k-1} A_{k-1} (\check{x}_{k-1} - \hat{x}_{k-1})\nonumber\\[2.25\jot]
	&\quad + (1-u_{k-1}) \imath_{k-1} + u_{k-1}(1-\gamma_{k-1}) \jmath_{k-1},
\end{align}
for $k \in \mathbb{N}_{[1,N]}$ with initial condition $\hat{x}_0 = m_0$, where $\imath_{k-1} = A_{k-1} \E[\hat{e}_{k-1} | \mathcal{I}_{k-1}^d, u_{k-1} = 0]$ and $\jmath_{k-1} = A_{k-1} \E[\hat{e}_{k-1} | \mathcal{I}_{k-1}^d, u_{k-1} = 1]$.
}
\end{lemma}

\begin{IEEEproof}
Observe that, given $\mathcal{I}^d_k$, the minimum mean-square-error estimator is $\E[ x_k | \mathcal{I}^d_k]$. Taking the conditional expectation of (\ref{eq:sys}) given $\mathcal{I}^d_k$, we obtain
\begin{align}\label{eq:propagation}
\E \Big[ x_k \big| \mathcal{I}^d_k \Big] = A_{k-1} \E \Big[x_{k-1} \big| \mathcal{I}_k^d \Big],
\end{align}
for $k \in \mathbb{N}_{[1,N]}$ as $w_{k-1}$ is independent of $\mathcal{I}^d_k$ and has zero mean. When sensory information is successfully delivered in the forward channel at time $k$, we have $z^f_k = \check{x}_{k-1}$. In this case, we get $\E[x_{k-1} | \mathcal{I}^d_k] = \E[x_{k-1} | \check{x}_{k-1}, Q_{k-1}] = \check{x}_{k-1}$ as $\{ \check{x}_{k-1}, Q_{k-1} \}$ is a sufficient statistic of $\mathcal{I}^d_k$. Hence, using (\ref{eq:propagation}), when $z^f_k = \check{x}_{k-1}$, i.e., $u_{k-1} = 1 \wedge \gamma_{k-1} = 1$, we~get
\begin{align}\label{eq:update0}
	\E \Big[x_k \big| \mathcal{I}^d_k \Big] = A_{k-1} \check{x}_{k-1},
\end{align}
for $k \in \mathbb{N}_{[1,N]}$. However, when no sensory information is delivered in the forward channel at time $k$, we have $z^f_k = \mathfrak{E}$ or $z^f_k = \mathfrak{F}$. Define $p_{k-1} := \E[x_{k-1} | \mathcal{I}^d_{k-1}, u_{k-1} = 0] - \E[x_{k-1} | \mathcal{I}^d_{k-1}]$ when $z^f_k = \mathfrak{E}$, and $q_{k-1} := \E[x_{k-1} | \mathcal{I}^d_{k-1}, u_{k-1} = 1] - \E[x_{k-1} | \mathcal{I}^d_{k-1}]$ when $z^f_k = \mathfrak{F}$. Then, using (\ref{eq:propagation}) and the definition of $p_{k-1}$, when $z^f_k = \mathfrak{E}$, i.e., $u_{k-1} = 0$, we get
\begin{align}\label{eq:erasure-update1}
	\E \Big[x_k \big| \mathcal{I}^d_k \Big] =  A_{k-1} \E \Big[x_{k-1} \big| \mathcal{I}^d_{k-1} \Big] + A_{k-1} p_{k-1},
\end{align}
for $k \in \mathbb{N}_{[1,N]}$, where we used the fact that $\{\mathcal{I}^d_{k-1}, z^f_k = \mathfrak{E}\}$ is a sufficient statistic of $\mathcal{I}^d_k$. Moreover, using (\ref{eq:propagation}) and the definition of $q_{k-1}$, when $z^f_k = \mathfrak{F}$, i.e., $u_{k-1} = 1 \wedge \gamma_{k-1} = 0$, we get
\begin{align}\label{eq:erasure-update2}
	\E \Big[x_k \big| \mathcal{I}^d_k \Big] =  A_{k-1} \E \Big[x_{k-1} \big| \mathcal{I}^d_{k-1} \Big] + A_{k-1} q_{k-1},
\end{align}
for $k \in \mathbb{N}_{[1,N]}$, where we used the facts that $\{\mathcal{I}^d_{k-1}, z^f_k = \mathfrak{F}\}$ is a sufficient statistic of $\mathcal{I}^d_k$ and that $x_{k-1}$ is independent of $\gamma_{k-1}$. Now, define $\imath_{k-1} := A_{k-1} p_{k-1}$ and $\jmath_{k-1} := A_{k-1} q_{k-1}$ as the signalling residuals. We can obtain (\ref{eq:est-monitor}) by combining (\ref{eq:update0}), (\ref{eq:erasure-update1}), and (\ref{eq:erasure-update2}). Finally, the initial condition is $\E[x_0] = m_0$, as no information is available at time $k = 0$.
\end{IEEEproof}

The next two lemmas provide certain properties pertaining to the estimation mismatch and the signalling residuals~$\imath_k$~and~$\jmath_k$.

\begin{lemma}\label{lem:mismatch-dyn-est}\emph{
The estimation mismatch satisfies the recursive equation
\begin{align}\label{eq:et-dynamics}
	\tilde{e}_k &= (1 - u_{k-1} \gamma_{k-1} ) A_{k-1} \tilde{e}_{k-1} + K_k \nu_k + \mu_{k-1},
\end{align}
for $k \in \mathbb{N}_{[1,N]}$ with initial condition $\tilde{e}_0 = K_0 \nu_0$, where $\mu_{k-1} = - (1-u_{k-1}) \imath_{k-1} - u_{k-1}(1-\gamma_{k-1}) \jmath_{k-1}$. Besides, the conditional mean and the conditional covariance of the estimation mismatch at the encoder satisfy the recursive equations
\begin{align}
	\breve{e}_k &= (1 - \bar{\theta}_{k-1} ) A_{k-1} \breve{e}_{k-1} + K_k \nu_k + \varsigma_{k-1}, \label{eq:lemma-eps-dyn}\\[2.25\jot]
	R_k &= (1 - \bar{\theta}_{k-1} ) A_{k-1} R_{k-1} A_{k-1}^T \nonumber\\[2.25\jot]
	&\quad + \bar{\bar{\theta}}_{k-1} A_{k-1} \breve{e}_{k-1} \breve{e}_{k-1}^T A_{k-1}^T + \Xi_{k-1}, \label{eq:lemma-R-dyn}
\end{align}
for $k \in \mathbb{N}_{[1,N]}$ with initial condition $\breve{e}_0 = K_0 \nu_0$ and $R_0 = 0$, where $\bar{\theta}_{k-1} = \theta_{k-1} u_{k-1} \gamma_{k-1} + \lambda^c (1 - \theta_{k-1}) u_{k-1}$, $\bar{\bar{\theta}}_{k-1} = (\lambda^c - {\lambda^c}^2) (1 - \theta_{k-1} )  u_{k-1}$, $\varsigma_{k-1} = \E[ \mu_{k-1} | \mathcal{I}^e_{k-1}, u_{k-1}, u_{k-1} \gamma_{k-1} = 0]$, and $\Xi_{k-1} = \Cov[ \mu_{k-1} | \mathcal{I}^e_{k-1}, u_{k-1}, u_{k-1} \gamma_{k-1} =0] + 2\Cov[\mu_{k-1}, A_{k-1} \tilde{e}_{k-1} | \mathcal{I}^e_{k-1}, u_{k-1}, u_{k-1} \gamma_{k-1} =0]$.}
\end{lemma}

\begin{IEEEproof}
Observe that we can obtain (\ref{eq:et-dynamics}) by plugging (\ref{eq:est-KF-m}) into (\ref{eq:est-KF-xc}), and then subtracting (\ref{eq:est-monitor}) from the result.

Suppose for now that $\mu_{k-1} = 0$. When feedback information is successfully delivered in the backward channel at time $k$, we have $z^b_k = \gamma_{k-1}$. Taking the conditional expectation and the conditional covariance of (\ref{eq:et-dynamics}) given $\mathcal{I}^e_k$, when $z^b_k = \gamma_{k-1}$, i.e., $u_{k-1} = 1 \wedge \theta_{k-1} = 1$, we get
\begin{align}
	\E \Big [\tilde{e}_k \Big| \mathcal{I}^e_k \Big]  &= (1\!-\!\sigma_{k-1} ) A_{k-1} \E \Big [ \tilde{e}_{k-1} \big| \mathcal{I}^e_{k-1} \Big]\!+\!K_k \nu_k, \label{eq:lem-eps-dyn-withgamma}\\[1\jot]
	\Cov \Big [\tilde{e}_k \Big| \mathcal{I}^e_k \Big] &= (1\!-\!\sigma_{k-1}) A_{k-1} \Cov \Big [\tilde{e}_{k-1} \big| \mathcal{I}^e_{k-1} \Big] A_{k-1}^T, \label{eq:lem-R-dyn-withgamma}
\end{align}
where $\sigma_{k-1} = u_{k-1} \gamma_{k-1}$, and we used the facts that $\{\mathcal{I}^e_{k-1}, y_k, z^b_k = \gamma_{k-1} \}$ is a sufficient statistic of $\mathcal{I}^e_k$, that $\tilde{e}_{k-1}$ is independent of $y_k$ and $z^b_k$ given $\mathcal{I}^e_{k-1}$, and that $\nu_k$ is known given $\mathcal{I}^e_k$. However, when no feedback information is delivered in the backward channel at time $k$, we have $z^b_k = \mathfrak{G}$ or $z^b_k = \mathfrak{H}$. Taking the conditional expectation and the conditional covariance of (\ref{eq:et-dynamics}) given $\mathcal{I}^e_k$, when $z^b_k = \mathfrak{G}$ or $z^b_k = \mathfrak{H}$, i.e., $u_{k-1} = 0$ or $u_{k-1} = 1 \wedge \theta_{k-1}=0$, respectively, we get
\begin{align}
	\E \Big [\tilde{e}_k \big| \mathcal{I}^e_k \Big]  &= (1\!-\!\lambda^c u_{k-1}) A_{k-1} \E \Big [ \tilde{e}_{k-1} \big| \mathcal{I}^e_{k-1} \Big] \nonumber \\[1\jot]
	&\ \ + K_k \nu_k, \label{eq:lem-eps-dyn-nogamma}\\[1\jot]
	\Cov \Big [\tilde{e}_k \big| \mathcal{I}^e_k \Big] &= (1 - \lambda^c u_{k-1}) A_{k-1} \Cov \Big [\tilde{e}_{k-1} \big| \mathcal{I}^e_{k-1} \Big] A_{k-1}^T \nonumber\\[1\jot]
	&\ \ + (\lambda^c u_{k-1} - {\lambda^c}^2 u_{k-1}) A_{k-1} \nonumber\\[1\jot]
	&\ \ \times \E \Big [ \tilde{e}_{k-1} \big| \mathcal{I}^e_{k-1} \Big] \E \Big [ \tilde{e}_{k-1} \Big| \mathcal{I}^e_{k-1} \Big]^T A_{k-1}^T, \label{eq:lem-R-dyn-nogamma}
\end{align}
where we used the facts  that $ \{\mathcal{I}^e_{k-1}, y_k, z^b_k = \mathfrak{G} \}$ or $\{\mathcal{I}^e_{k-1}, y_k, z^b_k = \mathfrak{H} \}$ is a sufficient statistic of $\mathcal{I}^e_k$, that $\gamma_{k-1}$ is independent of $\tilde{e}_{k-1}$ and $\mathcal{I}^e_k$, that $\tilde{e}_{k-1}$ is independent of $y_k$ and $z^b_k$ given $\mathcal{I}^e_{k-1}$, and that $\nu_k$ is known given $\mathcal{I}^e_k$, and the properties of covariances. Now, if $\mu_{k-1} \neq 0$, which happens only if $u_{k-1} \gamma_{k-1} = 0$, it is easy to see that the contributions of $\mu_{k-1}$ in $\E[\tilde{e}_k | \mathcal{I}^e_k ]$ and $\Cov[\tilde{e}_k | \mathcal{I}^e_k ]$ are captured by $\varsigma_{k-1} = \E[ \mu_{k-1} | \mathcal{I}^e_{k-1}, u_{k-1}, u_{k-1} \gamma_{k-1} =0]$ and $\Xi_{k-1} = \Cov[ \mu_{k-1} | \mathcal{I}^e_{k-1}, u_{k-1}, u_{k-1} \gamma_{k-1} =0] + 2\Cov[\mu_{k-1}, A_{k-1} \tilde{e}_{k-1} | \mathcal{I}^e_{k-1}, u_{k-1}, u_{k-1} \gamma_{k-1} =0]$, respectively. As a result, following the definitions of $\breve{e}_k$ and $R_k$, we can obtain (\ref{eq:lemma-eps-dyn}) by combining (\ref{eq:lem-eps-dyn-withgamma}) and (\ref{eq:lem-eps-dyn-nogamma}) and incorporating $\varsigma_{k-1}$, and can obtain~(\ref{eq:lemma-R-dyn}) by combining (\ref{eq:lem-R-dyn-withgamma}) and (\ref{eq:lem-R-dyn-nogamma}) and incorporating $\Xi_{k-1}$.
\end{IEEEproof}

\begin{lemma}\label{lem:zero-residuals}\emph{
Let $\Prob(u_k | \boldsymbol{\nu}_{0:k}, \boldsymbol{u}_{0:k-1})$ be a symmetric function of $\boldsymbol{\nu}_{0:k}$ for $k \in \mathbb{N}_{[0,N]}$. Then, $\imath_k$, $\jmath_k$, $\mu_k$, $\varsigma_k$, and $\Xi_k$ in Lemmas~\ref{lem:estimator-decoder} and \ref{lem:mismatch-dyn-est}, are equal to zero for $k \in \mathbb{N}_{[0,N-1]}$.}
\end{lemma}

\begin{IEEEproof}
Note that, by definition, $\mu_k$, $\varsigma_k$, and $\Xi_k$ are zero if $\imath_k$ and $\jmath_k$ are zero. Thus, we only need to prove the claim for $\imath_k$ and $\jmath_k$ for $k \in \mathbb{N}_{[0,N-1]}$. Observe that the initial condition $\tilde{e}_0$ is independent of the scheduling policy. We assume that $\imath_t = \jmath_t = 0$ for all $t \in \mathbb{N}_{[0,k-1]}$, and will show that $\imath_k = \jmath_k = 0$. We can write 
\begin{align*}
	&\Prob \Big( \boldsymbol{u}_{0:t} \big| \boldsymbol{\nu}_{0:t} \Big) = \prod_{t'=0}^{t} \Prob \Big(u_{t'} \big| \boldsymbol{\nu}_{0:t'}, \boldsymbol{u}_{0:t'-1} \Big),
\end{align*}
where we used the fact that $u_{t'}$ is independent of $\boldsymbol{\nu}_{t'+1:t}$ given $\boldsymbol{\nu}_{0:t'}$ and $\boldsymbol{u}_{0:t'-1}$. Therefore, by the hypothesis, we can deduce that $\Prob( \boldsymbol{u}_{0:t} | \boldsymbol{\nu}_{0:t})$ is symmetric with respect to $\boldsymbol{\nu}_{0:t}$. It follows that
\begin{align*}
	&\Prob \Big(\boldsymbol{\nu}_{0:t} \big| \boldsymbol{u}_{0:t}  \Big) \propto \Prob \Big( \boldsymbol{u}_{0:t} \big| \boldsymbol{\nu}_{0:t} \Big) \Prob \Big(\boldsymbol{\nu}_{0:t} \Big).
\end{align*}
Since $\Prob(\boldsymbol{\nu}_{0:t})$ is a symmetric distribution, we deduce that $\Prob(\boldsymbol{\nu}_{0:t} | \boldsymbol{u}_{0:t})$ is also symmetric with respect to $\boldsymbol{\nu}_{0:t}$. For any $s \in \mathbb{N}_{[0,t]}$, by marginalisation and the fact that $\boldsymbol{\nu}_{s:t}$ is independent of $\boldsymbol{u}_{0:s-1}$, we can find that $\Prob( \boldsymbol{\nu}_{s:t} | \boldsymbol{u}_{s:t})$ is symmetric with respect to $\boldsymbol{\nu}_{s:t}$.

Note that $z^f_{k+1} = \mathfrak{E}$ and $z^f_{k+1} = \mathfrak{F}$ are equivalent to $u_{k} = 0$ and $\{u_{k} = 1, \gamma_{k} = 0\}$, respectively. Accordingly, when $z^f_{k+1} = \mathfrak{E}$ or $z^f_{k+1} = \mathfrak{F}$, we can write
\begin{align*}
	&\E \bigg[ \hat{e}_k \Big| \mathcal{I}^d_k, z^f_{k+1} \bigg]\\
	&\qquad = \E \bigg[ \E \Big[ \hat{e}_k \big| \mathcal{I}^e_k, z^f_{k+1}, \boldsymbol{\gamma}_{0:k-1} \Big] \Big| \mathcal{I}^d_k, z^f_{k+1} \bigg]\\
	&\qquad= \E \bigg[ \E \Big[ \hat{e}_k \big| \mathcal{I}^e_k, \boldsymbol{\gamma}_{0:k-1} \Big] \Big| \mathcal{I}^d_k, z^f_{k+1} \bigg]\\
	&\qquad = \E \bigg[ \tilde{e}_k \Big| \mathcal{I}^d_k, z^f_{k+1} \bigg],
\end{align*}
where in the first equality we used the tower property of conditional expectations, and in the second equality the facts that $\hat{e}_k$ is independent of $\gamma_k$, and that $u_k \in \mathcal{I}^e_k$. In addition, from (\ref{eq:et-dynamics}), when $\imath_t = \jmath_t = 0$ for all $t \in \mathbb{N}_{[d-1,k-1]}$, we can write
\begin{align}\label{eq:lem6:et-dyn}
	\tilde{e}_k &= \sum_{t=1}^{\eta_k+1} \bigg(\prod_{t'=1}^{t-1} A_{k-t'}\bigg) K_{k-t+1} \nu_{k-t+1} \nonumber\\
	&= D_k \boldsymbol{\nu}_{k-\eta_k:k},
\end{align}
for $k \in \mathbb{N}_{[0,N]}$, where $\eta_k \in \mathbb{N}_{[0,k]}$ denotes the time elapsed since the last successful delivery when we are at time $k$ with the convention $\eta_k = k$ if no successful delivery has occurred, and $D_k$ is a matrix of proper dimension.

Clearly, at time $k$, the decoder knows exactly when the last successful delivery occurred, i.e., it knows $\eta_k$. Moreover, from the definition of $\eta_k$, when we are at time $k$, no successful delivery occurred from $k-\eta_k+1$ to $k$.

Let us define $a_t, b_t \in \{0,1\}$ such that $a_t b_t = 0$ for $t \in \mathbb{N}_{[0,N]}$. By (\ref{eq:lem6:et-dyn}) and the definitions of $\imath_k$ and $\jmath_k$, we obtain
\begin{align*}
	\imath_k &=  A_k D_k \E \bigg[  \boldsymbol{\nu}_{k-\eta_k:k} \Big| \mathcal{I}^d_{k}, z^f_{k+1}= \mathfrak{E}  \bigg]\\
			 &=A_k D_k \E \bigg[  \boldsymbol{\nu}_{k-\eta_k:k} \Big| \mathcal{I}^1_{k+1}  \bigg],\\[0\jot]
	\jmath_k &=  A_k D_k \E \bigg[ \boldsymbol{\nu}_{k-\eta_k:k} \Big| \mathcal{I}^d_{k}, z^f_{k+1} = \mathfrak{F} \bigg]\\
			 &=A_k D_k \E \bigg[ \boldsymbol{\nu}_{k-\eta_k:k} \Big| \mathcal{I}^2_{k+1}  \bigg],
\end{align*}
where $\mathcal{I}^1_{k+1} = \{ \mathcal{I}^d_{k-\eta_k}, \boldsymbol{u}_{k-\eta_k:k-1} = \boldsymbol{a}_{k-\eta_k:k-1}, u_k = 0, \boldsymbol{\gamma}_{k-\eta_k:k-1} = \boldsymbol{b}_{k-\eta_k:k-1}\}$ and $\mathcal{I}^2_{k+1} = \{ \mathcal{I}^d_{k-\eta_k}, \boldsymbol{u}_{k-\eta_k:k-1} = \boldsymbol{a}_{k-\eta_k:k-1}, u_k = 1, \boldsymbol{\gamma}_{k-\eta_k:k-1} = \boldsymbol{b}_{k-\eta_k:k-1}\}$, and we used the fact that $\boldsymbol{\nu}_{k-\eta_k:k}$ is independent of $\gamma_k$. Equivalently, we can write
\begin{align*}
	\imath_k &=  A_k D_k \E \bigg[  \boldsymbol{\nu}_{k-\eta_k:k} \Big| \bar{\mathcal{I}}^1_{k+1}  \bigg], \\[0\jot]
	\jmath_k &=  A_k D_k \E \bigg[ \boldsymbol{\nu}_{k-\eta_k:k} \Big| \bar{\mathcal{I}}^2_{k+1}  \bigg],
\end{align*}
where $\bar{\mathcal{I}}^1_{k+1} = \{ \boldsymbol{u}_{k-\eta_k:k-1} = \boldsymbol{a}_{k-\eta_k:k-1}, u_k = 0 \}$ and $\bar{\mathcal{I}}^2_{k+1} = \{ \boldsymbol{u}_{k-\eta_k:k-1} = \boldsymbol{a}_{k-\eta_k:k-1}, u_k = 1 \}$, and we used the fact that $\boldsymbol{\nu}_{k-\eta_k:k}$ is independent of $\mathcal{I}^d_{k-\eta_k}$ and $\boldsymbol{\gamma}_{k-\eta_k:k-1}$. We have already shown that $\Prob( \boldsymbol{\nu}_{k-\eta_k:k} | \boldsymbol{u}_{k-\eta_k:k})$ is symmetric with respect to $\boldsymbol{\nu}_{k-\eta_k-d+1:k}$. This implies that $\imath_k = 0$ and $\jmath_k = 0$, and consequently completes the proof.
\end{IEEEproof}

The next two technical lemmas are related to the symmetric decreasing rearrangements of non-negative functions. For the proofs of these lemmas, see, e.g., \cite{brock2000} and \cite{alvino1991}.

\begin{lemma}[Hardy-Littlewood inequality]\label{lemma:GHL}\emph{
Let $f$ and $g$ be non-negative functions defined on $\mathbb{R}^n$ that vanish at infinity. Then,
\begin{align}
	\int_{\mathbb{R}^n} f(x) g(x) dx \leq \int_{\mathbb{R}^n} f^*(x) g^*(x) dx,
\end{align}
where $f^*(x)$ and $g^*(x)$ are the symmetric decreasing rearrangements of $f(x)$ and $g(x)$.
}
\end{lemma}
\vspace{2mm}

\begin{lemma}\label{lemma:major}\emph{
Let $\mathcal{B}(r) \subseteq \mathbb{R}^n$ be a ball of radius $r$ centred at the origin, and $f$ and $g$ be non-negative functions defined on $\mathbb{R}^n$ that vanish at infinity and satisfy
\begin{align}
	\int_{\mathcal{B}(r)} f^*(x) dx \leq \int_{\mathcal{B}(r)} g^*(x) dx,
\end{align}
for all $r \geq 0$. Then,
\begin{align}
	\int_{\mathcal{B}(r)} h(x) f^*(x) dx \leq \int_{\mathcal{B}(r)} h(x) g^*(x) dx,
\end{align}
for all $r \geq 0$ and any non-negative non-increasing function $h$.}
\end{lemma}

The next lemma provides a loss function that is equivalent to the one under study.

\begin{lemma}\label{lem:equiv-loss}
Let $\tilde{e}_{0}$ be equal to $K_0 \nu_0$ and $\delta$ be constructed based on $\E[x_k | \mathcal{I}^d_k]$. Then, optimising the loss function $\Phi(\epsilon,\delta)$ over $\epsilon \in \mathcal{E}$ is equivalent to optimising the loss function
\begin{align}\label{eq:lemma:PSI}
	&\Psi^{N}_{\epsilon}(\tilde{e}_{0}) = \sum_{k=0}^{N} \Big\{ \alpha_{k} \Prob_{\epsilon}\Big(\boldsymbol{\sigma}_{0:k-1} = 0 \Big) \nonumber\\[0.5\jot]
	&\times  \E_{\epsilon} \Big[ u_{k} \big| \boldsymbol{\sigma}_{0:k-1} = 0 \Big] \nonumber\\[1\jot]
	& + \Prob_{\epsilon} \Big(\boldsymbol{\sigma}_{0:k-1} = 0 \Big) \E_{\epsilon} \Big[ \tilde{e}_{k}^T \tilde{e}_{k} \big| \boldsymbol{\sigma}_{0:k-1} = 0 \Big] \nonumber\\[1\jot]	
	& + \Prob_{\epsilon} \Big(\boldsymbol{\sigma}_{0:k-1} = 0, \sigma_{k} = 1 \Big) \nonumber \\[1\jot]
	&\times \E_{\epsilon} \Big[ \Psi^{k+1,N}_{\epsilon}(\tilde{e}_{k+1}) \big| \boldsymbol{\sigma}_{0:k-1} = 0, \sigma_{k} = 1 \Big] \Big\},
\end{align}
over $\epsilon \in \mathcal{E}$, where $\sigma_k = u_k \gamma_k$ and
\begin{align*}
	\Psi^{k,N}_{\epsilon}(\tilde{e}_k) =  \sum_{t=k}^{N} \E \Big[ \alpha_{t} u_{t} + \tilde{e}_{t}^T \tilde{e}_{t} \Big],
\end{align*}
for $k \in \mathbb{N}_{[1,N]}$ given $\tilde{e}_k = K_k \nu_k$.	
\end{lemma}

\begin{IEEEproof}
From the definition of the loss function $\Phi(\epsilon,\delta)$, we can write
\begin{align}\label{eq:Phi-lemma8}
	\Phi(\epsilon,\delta) &= \sum_{k=0}^{N} \E \bigg[ \alpha_k u_k + \hat{e}_k^T \hat{e}_k \bigg] \nonumber\\[0.5\jot]
	& = \sum_{k=0}^{N} \E \bigg[ \E \Big[\alpha_k u_k + \hat{e}_k^T \hat{e}_k \big| \mathcal{I}^e_k, \boldsymbol{\gamma}_{0:k-1} \Big] \bigg] \nonumber\\[0.5\jot]
	& = \sum_{k=0}^{N} \E \bigg[ \alpha_k u_k + \tilde{e}_k^T \tilde{e}_k + \tr Q_k \bigg],
\end{align}
where in the second equality we used the tower property of conditional expectations. Observe that, by Lemma~\ref{lem:mismatch-dyn-est}, we have $\tilde{e}_{k+1} = K_{k+1} \nu_{k+1}$ only when $\sigma_{k}=1$. We can show that, associated with the optimal scheduling policy, the following relation is satisfied:
\begin{align}\label{eq:propertyXXX}
	&\E \bigg[ \alpha_t u_t + \tilde{e}_t^T \tilde{e}_t \Big| \sigma_k = 1 \bigg]  \nonumber\\[0.75\jot]
	&\quad = \E \bigg[ \alpha_t u_{t} + \tilde{e}_t^T \tilde{e}_t \Big| \boldsymbol{\sigma}_{0:k-1}, \sigma_k = 1 \bigg],
\end{align}
for any $t\in \mathbb{N}_{[k+1,N]}$, where we used the fact that, without loss of optimality, $\boldsymbol{u}_{k+1:N}$ and $\tilde{\boldsymbol{e}}_{k+1:N}$ are independent of $\boldsymbol{\sigma}_{0:k-1}$ given $\tilde{e}_{k+1} = K_{k+1} \nu_{k+1}$.

Now, define the loss function $\Psi^{N}_{\epsilon}(\tilde{e}_0)$~as
\begin{align}\label{eq:Omega-lemma8}
	\Psi^{N}_{\epsilon}(\tilde{e}_0) := \sum_{k=0}^{N} \E \bigg[ \alpha_k u_k + \tilde{e}_k^T  \tilde{e}_k \bigg].
\end{align}
Following the facts that $\tilde{e}_0$ and $\tr Q_t$ for $t \in \mathbb{N}_{[0,N]}$ are independent of the scheduling policy, to optimise $\Phi(\epsilon,\delta)$ in (\ref{eq:Phi-lemma8}) over $\epsilon \in \mathcal{E}$, it suffices to optimise $\Psi^{N}_{\epsilon}(\tilde{e}_0)$ in (\ref{eq:Omega-lemma8}) over $\epsilon \in \mathcal{E}$. In addition, observe that from the law of total probability, we~have
\begin{align}\label{eq:identity2X}
	&\Prob_{\epsilon} \Big(\sigma_0 = 1\Big) + \Prob_{\epsilon}\Big(\boldsymbol{\sigma}_{0:t} = 0\Big) \nonumber\\[1\jot]
	& \qquad \qquad  + \sum_{t'=1}^{t} \Prob_{\epsilon}\Big(\boldsymbol{\sigma}_{0:t'-1} = 0, \sigma_{t'} = 1\Big) = 1,
\end{align}
for any $t \in \mathbb{N}_{[0,N]}$. Applying the law of total expectation for the terms $\E[ \alpha_{k} u_{k}]$ and $\E[\tilde{e}_{k}^T \tilde{e}_{k}]$ in $\Psi^N_{\epsilon}(\tilde{e}_{0})$ on a partition provided by the identity (\ref{eq:identity2X}) for $t = k-1$, repeating this procedure for $k \in \mathbb{N}_{[1,N]}$, and using the definition of $\Psi^{k,N}_{\epsilon}(\tilde{e}_k)$, we can obtain (\ref{eq:lemma:PSI}).
\end{IEEEproof}

We are now in a position to provide the proof of Theorem~\ref{thm:1}.

\begin{IEEEproof}
Suppose that $(\epsilon^o,\delta^o)$ is a coding policy profile that belongs to the set of globally optimal solutions, which cannot be empty. We prove global optimality of the policy profile $(\epsilon^\star,\delta^\star)$ in the claim by showing that $\Phi(\epsilon^\star,\delta^\star)$ cannot be greater than $\Phi^o:= \Phi(\epsilon^o,\delta^o)$. Without loss of generality, we assume that $m_0 = 0$. Similar arguments can be made for $m_0 \neq 0$ following a coordinate transformation. The proof consists of three steps.

\underline{\it{Step 1.}} We say a scheduling policy is innovation-based if, at each time $k$, it depends on $\boldsymbol{\nu}_{0:k}$ instead of $\boldsymbol{y}_{0:k}$, i.e., if it can be expressed in the form $\Prob(u_k | \boldsymbol{\nu}_{0:k}, \boldsymbol{z}^b_{0:k}, \boldsymbol{u}_{0:k-1}, \boldsymbol{\theta}_{0:k-1})$. In the first step of our proof, we show that $\Phi(\epsilon^n,\delta^o) = \Phi(\epsilon^o,\delta^o)$, where $\epsilon^n$ is an innovation-based scheduling policy. From the definition of the innovation, (\ref{eq:est-KF-xc}), and (\ref{eq:est-KF-m}), we can write 
\begin{align*}
\boldsymbol{y}_{0:k} &= \boldsymbol{\nu}_{0:k} + G_k \check{\boldsymbol{x}}_{0:k-1},\\[1.5\jot]
\check{\boldsymbol{x}}_{0:k} &= H_k \boldsymbol{\nu}_{0:k},
\end{align*}
where $G_k$ and $H_k$ are matrices of proper dimensions. Putting these equations together, we find $\boldsymbol{y}_{0:k} = \boldsymbol{\nu}_{0:k} + G_k H_{k-1} \boldsymbol{\nu}_{0:k-1}$. Therefore, $\Prob_{\epsilon^o}(u_k | \boldsymbol{y}_{0:k}, \boldsymbol{z}^b_{0:k}, \boldsymbol{u}_{0:k-1}, \boldsymbol{\theta}_{0:k-1})$ can equivalently be written as $\Prob_{\epsilon^n}(u_k | \boldsymbol{\nu}_{0:k}, \boldsymbol{z}^b_{0:k}, \boldsymbol{u}_{0:k-1}, \boldsymbol{\theta}_{0:k-1})$. This establishes that $\Phi(\epsilon^n,\delta^o) = \Phi(\epsilon^o,\delta^o)$. As our subsequent analysis is valid for any values of $\boldsymbol{z}^b_{0:k}$ and $\boldsymbol{\theta}_{0:k-1}$, for brevity, hereafter we omit the dependency of $\epsilon^n$ on these variables.

\underline{\it{Step 2.}} Let $\mathcal{B}(r)$ be a ball of radius $r$ centred at the origin and of proper dimension. Define $\varpi_k := T_k \boldsymbol{\nu}_{0:k} \in \mathcal{N}$ for a specific matrix $T_k$, and $\sigma_k := u_k \gamma_k$, $l_{\epsilon^s}(\varpi_k) := (\lambda + \lambda^c \Prob_{\epsilon^s}(u_k = 0| \varpi_k, \boldsymbol{\sigma}_{0:k-1} = 0)) \Prob_{\epsilon^s}( \varpi_k | \boldsymbol{\sigma}_{0:k-1} = 0)$, and $l_{\epsilon^n}(\varpi_k) := (\lambda + \lambda^c \Prob_{\epsilon^n}(u_k = 0| \varpi_k, \boldsymbol{\sigma}_{0:k-1} = 0)) \Prob_{\epsilon^n}( \varpi_k | \boldsymbol{\sigma}_{0:k-1} = 0)$. In the second step of our proof, we show that $\Phi(\epsilon^s,\delta^o) \leq \Phi(\epsilon^n,\delta^o)$, where $\epsilon^s$ is a special form of $\epsilon^n$ that is symmetric with respect to $\boldsymbol{\nu}_{0:k}$ at each time $k$. Let us construct $\epsilon^s$ such that the following condition is satisfied:
\begin{align}\label{eq:partII-construction}
\int_{\mathcal{B}(r)} l_{\epsilon^s}(\varpi_k) d \varpi_k = \int_{\mathcal{B}(r)} l^*_{\epsilon^n}(\varpi_k) d \varpi_k,
\end{align}
for all $r \geq 0$ with $l_{\epsilon^s}(\varpi_k)$ as a radially symmetric function of $\varpi_k$. We know that $\sigma_k = 0$ either when $u_k = 1 \; \wedge \; \gamma_k = 0$ or when $u_k = 0$ regardless of $\gamma_k$. Accordingly, we can write
\begin{align*}
	&\Prob_{\epsilon^n} \Big(\sigma_k = 0 \big| \varpi_k, \boldsymbol{\sigma}_{0:k-1} = 0 \Big) \\[1\jot]
	&\  = \Prob_{\epsilon^n} \Big(u_k = 1 \big| \varpi_k, \boldsymbol{\sigma}_{0:k-1} = 0 ) \\[1\jot]
	&\qquad \times \Prob_{\epsilon^n} \Big(u_k \gamma_k = 0 \big| \varpi_k, \boldsymbol{\sigma}_{0:k-1} = 0, u_k = 1 \Big) \\[1\jot]
	&\qquad + \Prob_{\epsilon^n} \Big(u_k = 0 \big| \varpi_k, \boldsymbol{\sigma}_{0:k-1} = 0 \Big) \\[1\jot]
	&\qquad \times  \Prob_{\epsilon^n} \Big(u_k \gamma_k = 0 \big| \varpi_k, \boldsymbol{\sigma}_{0:k-1} = 0, u_k = 0 \Big)
\end{align*}
\begin{align*}
	&\ = \lambda \Prob_{\epsilon^n} \Big( u_k = 1 \big| \varpi_k, \boldsymbol{\sigma}_{0:k-1} = 0 \Big) \\[1\jot]
	&\qquad + \Prob_{\epsilon^n} \Big(u_k = 0 \big| \varpi_k, \boldsymbol{\sigma}_{0:k-1} = 0 \Big)\\[1\jot]
	&\ = \lambda + \lambda^c \Prob_{\epsilon^n} \Big(u_k = 0 \big| \varpi_k , \boldsymbol{\sigma}_{0:k-1} = 0 \Big),
\end{align*}
where in the first equality we used the law of total probability and in the second equality the fact that $\gamma_k$ is independent of $\varpi_k$, $\boldsymbol{\sigma}_{0:k-1}$, and $u_k$. Define $l'_{\epsilon^s}(\varpi_k) = \Prob_{\epsilon^s}(\sigma_k = 0| \varpi_k, \boldsymbol{\sigma}_{0:k-1} = 0) \Prob_{\epsilon^s}( \varpi_k | \boldsymbol{\sigma}_{0:k-1} = 0)$ and $l'_{\epsilon^s}(\varpi_k) = \Prob_{\epsilon^s}(\sigma_k = 0| \varpi_k, \boldsymbol{\sigma}_{0:k-1} = 0) \Prob_{\epsilon^s}( \varpi_k | \boldsymbol{\sigma}_{0:k-1} = 0)$. Therefore, the above condition can be written in the following equivalent form:
\begin{align}\label{eq:partII-construction-equiv}
\int_{\mathcal{B}(r)} l'_{\epsilon^s}(\varpi_k) d \varpi_k = \int_{\mathcal{B}(r)} {l'}^*_{\epsilon^n}(\varpi_k) d \varpi_k,
\end{align}
for all $r \geq 0$ with $l'_{\epsilon^s}(\varpi_k)$ as a radially symmetric function of $\varpi_k$. By Lemma~\ref{lem:equiv-loss}, to prove that $\Phi(\epsilon^s,\delta^o) \leq \Phi(\epsilon^n,\delta^o)$, it suffices to prove that $\Psi^{M}_{\epsilon^s}(\tilde{e}_{0}) \leq \Psi^M_{\epsilon^n}(\tilde{e}_{0})$ for any $M \in \mathbb{N}_{[0,N]}$ and for any $\tilde{e}_{0}$. Observe that $\tilde{e}_{0} = K_0 \nu_0$ is the same under both $\epsilon^n$ and $\epsilon^s$, and that $u_0$ has no effect on remote estimation when the time horizon is $0$. Hence, the claim holds for the time horizon $0$. We assume that the claim also holds for all time horizons from $1$ to $M-1$, and will show that the terms in $\Psi^{M}_{\epsilon^n}(\tilde{e}_{0})$ are not less than those in $\Psi^{M}_{\epsilon^s}(\tilde{e}_{0})$.

\underline{\it{Step 2.1}} First, for the probability coefficients, we have
\begin{align*}
&\Prob_{\epsilon^n} \Big(\sigma_{k-1} = 0 \big| \boldsymbol{\sigma}_{0:k-2} = 0 \Big)\\[1\jot]
&\ = \int_{\mathbb{R}^m} \Prob_{\epsilon^n}(\sigma_{k-1} = 0| \varpi_{k-1}, \boldsymbol{\sigma}_{0:k-2} = 0)\\[1.5\jot]
&\qquad \qquad \qquad \times \Prob_{\epsilon^n}(\varpi_{k-1} | \boldsymbol{\sigma}_{0:k-2} = 0) d \varpi_{k-1}\\[1.5\jot]
&\ = \int_{\mathbb{R}^m} \Prob_{\epsilon^s}(\sigma_{k-1} = 0| \varpi_{k-1}, \boldsymbol{\sigma}_{0:k-2} = 0)\\[1.5\jot]
&\qquad \qquad \qquad \times \Prob_{\epsilon^s}(\varpi_{k-1} | \boldsymbol{\sigma}_{0:k-2} = 0) d \varpi_{k-1}\\[1.5\jot]
&\ =\Prob_{\epsilon^s} \Big(\sigma_{k-1} = 0 \big| \boldsymbol{\sigma}_{0:k-2} = 0\Big),
\end{align*}
where the second equality comes from (\ref{eq:partII-construction-equiv}) when $r \to \infty$, following the fact that $\int_{\mathbb{R}^m} {l'}^*_{\epsilon^n}(\varpi_k) d \varpi_k = \int_{\mathbb{R}^m} l'_{\epsilon^n}(\varpi_k) d \varpi_k$. This also implies that $\Prob_{\epsilon^n}(\boldsymbol{\sigma}_{0:k-1} = 0) = \Prob_{\epsilon^s}(\boldsymbol{\sigma}_{0:k-1} = 0) $ and that $\Prob_{\epsilon^n}(\boldsymbol{\sigma}_{0:k-1} = 0, \sigma_{k} = 1) = \Prob_{\epsilon^s}(\boldsymbol{\sigma}_{0:k-1} = 0, \sigma_{k} = 1)$. 

\underline{\it{Step 2.2.}} Moreover, for the terms involving the expected value of the transmission decision, we have
\begin{align*}
\E_{\epsilon^n} \Big[ & u_{k} \big| \boldsymbol{\sigma}_{0:k-1} = 0 \Big]\\[2\jot]
	&= 1 - \Prob_{\epsilon^n}(u_{k} = 0| \boldsymbol{\sigma}_{0:k-1} = 0)\\[2\jot]
	&= \frac{1}{\lambda^c} - \frac{1}{\lambda^c} \Prob_{\epsilon^n}(\sigma_{k} = 0| \boldsymbol{\sigma}_{0:k-1} = 0)\\[2\jot]
	&= \frac{1}{\lambda^c} - \frac{1}{\lambda^c} \Prob_{\epsilon^s}(\sigma_{k} = 0| \boldsymbol{\sigma}_{0:k-1} = 0)\\[2\jot]
	&= 1 - \Prob_{\epsilon^s}(u_{k} = 0| \boldsymbol{\sigma}_{0:k-1} = 0)\\[2\jot]
	&=\E_{\epsilon^s} \Big[ u_{k} \big| \boldsymbol{\sigma}_{0:k-1} = 0 \Big].
\end{align*}

\underline{\it{Step 2.3.}} Observe that by Lemma~\ref{lem:mismatch-dyn-est}, when $\boldsymbol{\sigma}_{0:k-1}=0$, $\tilde{e}_t$ satisfies the recursive equation
\begin{align*}
	\tilde{e}_t &= A_{t-1} \tilde{e}_{t-1} + K_t \nu_t - (1-u_{t-1}) \imath_{t-1} - u_{t-1} \jmath_{t-1},
\end{align*}
for $t \in \mathbb{N}_{[1,k]}$ with initial condition $\tilde{e}_0 = K_0 \nu_0$. Accordingly, we can find a proper matrix $E_k$ and a proper vector $c_k$, both independent of $\boldsymbol{\nu}_{0:k}$, such that $\tilde{e}_k = E_k \boldsymbol{\nu}_{0:k-1} + K_k \nu_{k} + c_k$ under $\epsilon^n$. We know that $\epsilon^s$ is symmetric with respect to $\boldsymbol{\nu}_{0:k}$ at each time $k$. Therefore, by Lemma~\ref{lem:zero-residuals}, we deduce that $\tilde{e}_k = E_k \boldsymbol{\nu}_{0:k-1} + K_k \nu_{k}$ under $\epsilon^s$. For the terms involving the expected value of the quadratic estimation mismatch, we can then write
\begin{align*}
	&\E_{\epsilon^n} \Big[ \tilde{e}_k^T \tilde{e}_k \big| \boldsymbol{\sigma}_{0:k-1} = 0 \Big]\\[1.5\jot]
	&\ = \E_{\epsilon^n} \Big[ \big(E_k \boldsymbol{\nu}_{0:k-1} + K_k \nu_{k} + c_k \big)^T\\[1.5\jot]
	&\qquad \times \big(E_k \boldsymbol{\nu}_{0:k-1} + K_k \nu_{k} + c_k \big) \big| \boldsymbol{\sigma}_{0:k-1} = 0 \Big]\\[1.5\jot]
	&\ = \E_{\epsilon^n} \Big[ \big(E_k \boldsymbol{\nu}_{0:k-1} + c_k \big)^T \big(E_k \boldsymbol{\nu}_{0:k-1} + c_k \big)\\[2\jot]
	&\qquad + \nu_{k}^T K_k^T K_k \nu_{k} \big| \boldsymbol{\sigma}_{0:k-1} = 0 \Big],
\end{align*}
where in the second equality we used the fact that $\nu_{k}$ has zero mean and is independent of $\boldsymbol{\nu}_{0:k-1}$ and $\boldsymbol{\sigma}_{0:k-1}$. Choose $T_{k-1} = E_k$, and define $f_{\epsilon^n}(\varpi_{k-1}, \nu_{k}) := (\varpi_{k-1} + c_k)^T (\varpi_{k-1} + c_k) + \nu_{k}^T K_k^T K_k \nu_{k}$, $f_{\epsilon^s}(\varpi_{k-1}, \nu_{k}) := \varpi_{k-1}^T \varpi_{k-1} + \nu_{k}^T K_k^T K_k \nu_{k}$, $g_{\epsilon^n}(\varpi_{k-1}, \nu_{k}) := z -\min_z \{z,f_{\epsilon^n}(\varpi_{k-1}, \nu_{k})\}$, and $g_{\epsilon^s}(\varpi_{k-1}, \nu_{k}) := z -\min_z \{z,f_{\epsilon^s}(\varpi_{k-1}, \nu_{k})\}$. Clearly, $g_{\epsilon^n}(\varpi_{k-1}, \nu_{k})$ and $g_{\epsilon^s}(\varpi_{k-1}, \nu_{k})$ both vanish at infinity for any fixed~$z$. It follows~that
\begin{align*}
	&\E_{\epsilon^n} \Big[ \tilde{e}_k^T \tilde{e}_k \big| \boldsymbol{\sigma}_{0:k-1} = 0 \Big]\\[1.75\jot]
	&\qquad =  \int_{\mathbb{R}^m} \int_{\mathbb{R}^m}  f_{\epsilon^n}(\varpi_{k-1}, \nu_{k}) \Prob_{\epsilon^n}( \varpi_{k-1} | \boldsymbol{\sigma}_{0:k-1} = 0)\\[2.5\jot]
	&\qquad \qquad \qquad \qquad \times \Prob(\nu_{k}) d \varpi_{k-1} d \nu_{k} .
\end{align*}
In addition, we can write
\begin{align*}
	& \int_{\mathbb{R}^m} g_{\epsilon^n}(\varpi_{k-1}, \nu_{k}) \Prob_{\epsilon^n}( \varpi_{k-1} | \boldsymbol{\sigma}_{0:k-2} = 0)\\[1.75\jot]
	&\qquad \times \Prob_{\epsilon^n}(\sigma_{k-1} = 0 | \varpi_{k-1}, \boldsymbol{\sigma}_{0:k-2} = 0)  d \varpi_{k-1}\\[1.75\jot]
	&\leq  \int_{\mathbb{R}^m} g^*_{\epsilon^n}(\varpi_{k-1}, \nu_{k}) \big(\Prob_{\epsilon^n}( \varpi_{k-1} | \boldsymbol{\sigma}_{0:k-2} = 0)  \\[1.25\jot]
	&\qquad \times \Prob_{\epsilon^n}(\sigma_{k-1} = 0 | \varpi_{k-1}, \boldsymbol{\sigma}_{0:k-2} = 0) \big)^* d \varpi_{k-1}\\[1.75\jot]
	&= \int_{\mathbb{R}^m} g_{\epsilon^s}(\varpi_{k-1}, \nu_{k}) \big( \Prob_{\epsilon^n}( \varpi_{k-1} | \boldsymbol{\sigma}_{0:k-2} = 0) \\[1.25\jot]
	&\qquad \times \Prob_{\epsilon^n}(\sigma_{k-1} = 0 | \varpi_{k-1}, \boldsymbol{\sigma}_{0:k-2} = 0) \big)^* d \varpi_{k-1}\\[1.75\jot]
	&\leq \int_{\mathbb{R}^m} g_{\epsilon^s}(\varpi_{k-1}, \nu_{k}) \Prob_{\epsilon^s}( \varpi_{k-1} | \boldsymbol{\sigma}_{0:k-2} = 0) \\[1.75\jot]
	&\qquad \times \Prob_{\epsilon^s}(\sigma_{k-1} = 0 | \varpi_{k-1}, \boldsymbol{\sigma}_{0:k-2} = 0)  d \varpi_{k-1},
\end{align*}
where in the first inequality we used the Hardy-Littlewood inequality with respect to $\varpi_{k-1}$, in the equality the fact that $g^*_{\epsilon^n}(\varpi_{k-1}, \nu_{k}) = g_{\epsilon^s}(\varpi_{k-1}, \nu_{k})$, and in the second inequality~(\ref{eq:partII-construction-equiv}) and Lemma~\ref{lemma:major}. This implies that
\begin{align}\label{eq:min_z_and_f}
	& \int_{\mathbb{R}^m} \underset{z}{\min} \Big\{z,f_{\epsilon^n}(\varpi_{k-1}, \nu_{k}) \Big\} \nonumber\\[1.5\jot]
	&\qquad \qquad \qquad \times \Prob_{\epsilon^n}(\varpi_{k-1} | \boldsymbol{\sigma}_{0:k-1} = 0) d \varpi_{k-1} \nonumber\\[1.5\jot]
	&\geq \int_{\mathbb{R}^m} \underset{z}{\min} \Big \{z,f_{\epsilon^s}(\varpi_{k-1}, \nu_{k}) \Big\}\nonumber\\[1.5\jot]
	&\qquad \qquad \qquad  \times \Prob_{\epsilon^s}(\varpi_{k-1} | \boldsymbol{\sigma}_{0:k-1} = 0) d \varpi_{k-1},
\end{align}
where we applied the Bayes' rule on $\Prob_{\epsilon^n}(\varpi_{k-1} | \boldsymbol{\sigma}_{0:k-1} = 0)$ and used the fact that $\Prob_{\epsilon^n}(\sigma_{k-1} = 0|\boldsymbol{\sigma}_{0:k-2} = 0) = \Prob_{\epsilon^s}(\sigma_{k-1} = 0|\boldsymbol{\sigma}_{0:k-2} = 0)$. Then, we deduce that
\begin{align*}
	&\E_{\epsilon^n} \Big[ \tilde{e}_k^T \tilde{e}_k \big| \boldsymbol{\sigma}_{0:k-1} = 0 \Big]\\[2\jot]
	&\quad =  \int_{\mathbb{R}^m} \int_{\mathbb{R}^m}  f_{\epsilon^n}(\varpi_{k-1}, \nu_{k}) \Prob_{\epsilon^n}(\varpi_{k-1} | \boldsymbol{\sigma}_{0:k-1} = 0)\\[1.5\jot]
	&\qquad \qquad \qquad \qquad \times \Prob( \nu_{k}) d \varpi_{k-1} d \nu_k\\[1.5\jot]
	&\quad \geq \int_{\mathbb{R}^m} \int_{\mathbb{R}^m}  f_{\epsilon^s}(\varpi_{k-1}, \nu_{k}) \Prob_{\epsilon^s}(\varpi_{k-1} | \boldsymbol{\sigma}_{0:k-1} = 0)\\[1.5\jot]
	&\qquad \qquad \qquad \qquad \times  \Prob( \nu_{k}) d \varpi_{k-1} d \nu_k\\[1.5\jot]
	&\quad =\E_{\epsilon^s} \Big[ \tilde{e}_k^T \tilde{e}_k \big| \boldsymbol{\sigma}_{0:k-1} = 0 \Big],
\end{align*}
where the inequality is obtained from (\ref{eq:min_z_and_f}) after taking $z$ to infinity.

\underline{\it{Step 2.4.}} In addition, for the terms involving the expected value of the cost-to-go, we find 
\begin{align*}
	&\E_{\epsilon^n} \Big[ \Psi^{k+1,M}_{\epsilon^n}(\tilde{e}_{k+1}) \big| \boldsymbol{\sigma}_{0:k-1} = 0 , \sigma_{k} = 1 \Big]\\[1\jot]
	&\quad = \int_{\mathbb{R}^{m\times(k+2)}} \Psi^{k+1,M}_{\epsilon^n}(\tilde{e}_{k+1})\\[2\jot]
	&\qquad \qquad \times \Prob_{\epsilon^n}(\boldsymbol{\nu}_{0:k+1} | \boldsymbol{\sigma}_{0:k-1} = 0, \sigma_{k} = 1) d \boldsymbol{\nu}_{0:k+1}.
\end{align*}
By Lemma~\ref{lem:mismatch-dyn-est}, when $\sigma_{k}=1$, we have $\tilde{e}_{k+1} = K_{k+1} \nu_{k+1}$. Hence, $\tilde{e}_{k+1}$ is the same under both $\epsilon^n$ and $\epsilon^s$. For any $M$, let $\bar{\Psi}^{M}_{\epsilon^n}(\tilde{e}_{0})$ represent a loss function that is structurally similar to $\Psi^{M}_{\epsilon^n}(\tilde{e}_{0})$ but with new values of $\boldsymbol{\alpha}_{0:M}$, say $\bar{\boldsymbol{\alpha}}_{0:M}$. Clearly, if $\Psi^M_{\epsilon^n}(\tilde{e}_{0}) \geq \Psi^M_{\epsilon^s}(\tilde{e}_{0})$, then $\bar{\Psi}^M_{\epsilon^n}(\tilde{e}_{0}) \geq \bar{\Psi}^M_{\epsilon^s}(\tilde{e}_{0})$. Accordingly, we can write
\begin{align*}
	& \int_{\mathbb{R}^{m\times(k+2)}} \Psi^{k+1,M}_{\epsilon^n} (\tilde{e}_{k+1})\\[2\jot]
	&\qquad \quad \times \Prob_{\epsilon^n}(\boldsymbol{\nu}_{0:k+1} | \boldsymbol{\sigma}_{0:k-1}=0, \sigma_{k}=1) d \boldsymbol{\nu}_{0:k+1}\\[2\jot]
	&= \int_{\mathbb{R}^m} \bar{\Psi}^{M-k-1}_{\epsilon^n}(K_{k+1} \nu_{k+1}) \Prob(\nu_{k+1}) d \nu_{k+1}\\[1.75\jot]
	&\geq \int_{\mathbb{R}^m} \bar{\Psi}^{M-k-1}_{\epsilon^s}(K_{k+1} \nu_{k+1}) \Prob(\nu_{k+1}) d \nu_{k+1}\\[1.75\jot]
	&=  \int_{\mathbb{R}^{m\times(k+2)}} \Psi^{k+1,M}_{\epsilon^s}(\tilde{e}_{k+1})\\[2\jot]
	&\qquad \quad \times \Prob_{\epsilon^s}(\boldsymbol{\nu}_{0:k+1} | \boldsymbol{\sigma}_{0:k-1}=0, \sigma_{k}=1) d \boldsymbol{\nu}_{0:k+1},
\end{align*}
where in the equalities we used the facts that $\Psi^{k+1,M}_{\epsilon^n}(\tilde{e}) = \bar{\Psi}^{M-k-1}_{\epsilon^n}(\tilde{e})$ for any Gaussian variable $\tilde{e}$ and a suitable selection of $\bar{\boldsymbol{\alpha}}_{0:M-k-1}$, and that $\nu_{k+1}$ is independent of $\boldsymbol{\sigma}_{0:k}$, and the Fubini's theorem; and in the inequality we used the hypothesis $\Psi^{M-k-1}_{\epsilon^n}(\tilde{e}) \geq \Psi^{M-k-1}_{\epsilon^s}(\tilde{e})$ for any Gaussian variable $\tilde{e}$. Therefore,
\begin{align*}
&\E_{\epsilon^n} \Big[ \Psi^{k+1,M}_{\epsilon^n}(\tilde{e}_{k+1}) \big| \boldsymbol{\sigma}_{0:k-1} = 0 , \sigma_{k} = 1\Big]\\[1.5\jot]
&\qquad \geq \E_{\epsilon^s} \Big[ \Psi^{k+1,M}_{\epsilon^s}(\tilde{e}_{k+1}) \big| \boldsymbol{\sigma}_{0:k-1} = 0 , \sigma_{k} = 1\Big].
\end{align*}
This establishes that $\Psi^M_{\epsilon^s}(\tilde{e}_{0}) \leq \Psi^M_{\epsilon^n}(\tilde{e}_{0})$, and verifies that $\Phi(\epsilon^s,\delta^o) \leq \Phi(\epsilon^n,\delta^o)$.

\underline{\it{Step 3.}} Now, in the final step of our proof, we show that $\Phi(\epsilon^\star,\delta^\star) \leq \Phi(\epsilon^s,\delta^o)$, where $\epsilon^\star$ is a special form of $\epsilon^s$ and $\delta^\star$ is the same as $\delta^o$. Observe that, by Lemma~\ref{lemma0} and Lemmas~\ref{lem:estimator-decoder} and \ref{lem:zero-residuals}, for any $\epsilon^s$ that is adopted, $\delta^o$ must satisfy the recursive equation
\begin{align}\label{eq:proof-filter-sym}
	&\hat{x}_{k} =  A_{k-1} \hat{x}_{k-1} +  u_{k-1} \gamma_{k-1} A_{k-1} (\check{x}_{k-1} - \hat{x}_{k-1}),
\end{align}
for $k \in \mathbb{N}_{[1,N]}$ with initial condition $\hat{x}_0 = m_0 = 0$. Moreover, from the definition of the value function $V_k(\mathcal{I}^e_k)$ in (\ref{eq:Ve-def}), we can write
\begin{align}\label{eq:valuefunction-inproof}
	&V_{k}(\mathcal{I}^e_k) = \min_{u_k \in \{0, 1\} } \bigg\{\alpha_k u_k + \E[ \tilde{e}_{k+1}^T \tilde{e}_{k+1} | \mathcal{I}^e_k] \nonumber\\[0\jot]
	&\qquad \qquad \qquad \qquad + \tr Q_{k+1}  + \E[V_{k+1}(\mathcal{I}^e_{k+1})|\mathcal{I}^e_k]  \bigg\},
\end{align}
for $k \in \mathbb{N}_{[0,N-1]}$ with initial condition $V_{N}(\mathcal{I}^e_{N}) = 0$, where we used the additivity property of $V_{k}(\mathcal{I}^e_k)$ and the fact that $\E[ \hat{e}_{k+1}^T \hat{e}_{k+1} | \mathcal{I}^e_k] = \E[ \E[ \hat{e}_{k+1}^T \hat{e}_{k+1}| \boldsymbol{\gamma}_{0:k},\mathcal{I}^e_{k+1}] | \mathcal{I}^e_k] = \E[ \tilde{e}_{k+1}^T \tilde{e}_{k+1} | \mathcal{I}^e_k] + \tr Q_{k+1}$ according to the tower properties of conditional expectations. The minimising scheduling policy is then obtained~as
\begin{align}\label{eq:stage3-delta}
	u_k = \mathds{1}_{\chi_k - \alpha_k \geq0},
\end{align}
where $\chi_k = \E[ \tilde{e}_{k+1}^T \tilde{e}_{k+1} + V_{k+1}(\mathcal{I}^e_{k+1})| \mathcal{I}^e_k, u_k = 0] - \E[ \tilde{e}_{k+1}^T \tilde{e}_{k+1} + V_{k+1}(\mathcal{I}^e_{k+1})| \mathcal{I}^e_k,u_k = 1]$. Next, we prove by induction that $V_{k}(\mathcal{I}^e_k)$ is a symmetric function of $\breve{e}_k$ and a function $R_k$. Clearly, the claim holds for time $N$. We assume that the claim holds for time $k+1$ and we will show that it also holds for time $k$. Let us write $\chi_k = \chi^1_k + \chi^2_k$, where $\chi^1_k = \E[ \tilde{e}_{k+1}^T \tilde{e}_{k+1}| \mathcal{I}^e_k, u_k = 0] - \E[ \tilde{e}_{k+1}^T \tilde{e}_{k+1} | \mathcal{I}^e_k,u_k = 1]$ and $\chi^2_k = \E[ V_{k+1}(\breve{e}_{k+1}, R_{k+1})| \mathcal{I}^e_k, u_k = 0] - \E[ V_{k+1}(\breve{e}_{k+1}, R_{k+1})| \mathcal{I}^e_k,u_k = 1]$. From the properties of expectations, we have the following relation
\begin{align*}
	\E \Big[ \tilde{e}_{k+1}^T \tilde{e}_{k+1} \big| \mathcal{I}^e_k, u_k \Big] &= \E \Big[ \tilde{e}_{k+1} \big| \mathcal{I}^e_k, u_k \Big]^T \E \Big[ \tilde{e}_{k+1} \big| \mathcal{I}^e_k, u_k \Big]\\[2\jot]
	&\quad + \tr \Cov \Big[ \tilde{e}_{k+1} \big| \mathcal{I}^e_k, u_k \Big].
\end{align*}
By Lemmas~\ref{lem:mismatch-dyn-est} and \ref{lem:zero-residuals}, we can write
\begin{align*}
	\tilde{e}_{k+1} &= (1 - u_{k} \gamma_{k} ) A_{k} \tilde{e}_{k} + K_{k+1} \nu_{k+1}.
\end{align*}
Accordingly, the conditional mean and the conditional covariance of $\tilde{e}_{k+1}$ given $\mathcal{I}^e_k$ and $u_k$ are obtained as
\begin{align*}
	\E \Big[ \tilde{e}_{k+1} \big| \mathcal{I}^e_k, u_k \Big] &= (1 - \lambda^c u_k) A_k \breve{e}_k, \\[1\jot]
	\Cov \Big[ \tilde{e}_{k+1} \big| \mathcal{I}^e_k, u_k \Big] &= (\lambda^c u_k - {\lambda^c}^2 u_k) A_k \breve{e}_k \breve{e}_k^T A_k^T\\[2.25\jot]
	&\quad + (1 - \lambda^c u_k)A_k R_k A_k^T\\[2.5\jot]
	&\quad + K_{k+1} N_{k+1} K_{k+1}^T,
\end{align*}
where we used the fact that $\nu_{k+1}$ has zero mean and is independent of $\tilde{e}_k$ and $\gamma_k$, and the properties of covariances. Thus, we get
\begin{align*}
	\E \Big[ \tilde{e}_{k+1}^T \tilde{e}_{k+1} \big| \mathcal{I}^e_k, u_k \Big] &=  (1 - \lambda^c u_k) \breve{e}_k^T A_k^T A_k \breve{e}_k \\[2\jot]
	&\quad + (1 - \lambda^c u_k) \tr(A_k R_k A_k^T)\\[2.5\jot]
	&\quad + \tr(K_{k+1} N_{k+1} K_{k+1}^T).
\end{align*}
This implies that $\E [ \tilde{e}_{k+1}^T \tilde{e}_{k+1} | \mathcal{I}^e_k, u_k ]$ is a symmetric function of $\breve{e}_k$ and a function of $R_k$, and that
\begin{align*}
	\chi^1_k = \lambda^c \breve{e}_k^T A_k^T A_k \breve{e}_k + \lambda^c \tr(A_k R_k A_k^T).
\end{align*}	
Therefore, we deduce that $\chi^1_k$ is also a symmetric function of $\breve{e}_k$ and a function of $R_k$. Besides, by Lemmas~\ref{lem:mismatch-dyn-est} and \ref{lem:zero-residuals}, we can~write
\begin{align}
	\breve{e}_{k+1} &= (1 - \bar{\theta}_{k}) A_{k} \breve{e}_{k} + K_{k+1} \nu_{k+1}, \label{eq:step3-eps}\\[2.5\jot]
	R_{k+1} &= (1 - \bar{\theta}_{k}) A_{k} R_{k} A_{k}^T + \bar{\bar{\theta}}_{k} A_{k} \breve{e}_{k} \breve{e}_{k}^T A_{k}^T, \label{eq:step3-R}
\end{align}
where $\bar{\theta}_k = \theta_k u_k \gamma_k + \lambda^c (1 - \theta_k) u_k$ and $\bar{\bar{\theta}}_k = (\lambda^c - {\lambda^c}^2) (1 - \theta_k ) u_k$. By the hypothesis, $V_{k+1}(\breve{e}_{k+1}, R_{k+1})$ is a symmetric function of $\breve{e}_{k+1}$ and a function of $R_{k+1}$. Plugging (\ref{eq:step3-eps}) and (\ref{eq:step3-R}) into $V_{k+1}(\breve{e}_{k+1}, R_{k+1})$, we can calculate $\E[ V_{k+1}(\breve{e}_{k+1}, R_{k+1})| \mathcal{I}^e_k,u_k]$ while all $\nu_{k+1}$, $\theta_k$, and $\gamma_k$ are averaged out. A straightforward consequence is that we can express $\E[ V_{k+1}(\breve{e}_{k+1}, R_{k+1})| \mathcal{I}^e_k,u_k]$ as a symmetric function of $\breve{e}_k$ and a function of $R_k$, since $\nu_{k+1}$ is a Gaussian variable with zero mean. Therefore, we deduce that $\chi^2_k$ is also a symmetric function of $\breve{e}_k$ and a function $R_k$. Thus, using (\ref{eq:stage3-delta}), we conclude that the claim holds.
\end{IEEEproof}

\bibliography{mybib}
\bibliographystyle{ieeetr}

\newpage


\begin{IEEEbiographynophoto}{Touraj~Soleymani} received his Ph.D. degree in electrical and computer engineering from the Technical University of Munich, Germany, in 2019, and his M.S. and B.S. degrees both in aeronautical engineering from Sharif University of Technology, Iran, in 2011 and 2008, respectively. He is currently an assistant professor at the University of London, United Kingdom. He was a research associate in the Department of Electrical and Electronic Engineering, Imperial College London, United Kingdom, from 2022 to 2024; a research associate at the School of Electrical Engineering and Computer Science, Royal Institute of Technology, Sweden, from 2019 to 2022; a research assistant at the Institute for Advanced Study, Technical University of Munich, Germany, from 2014 to 2019; and a research assistant at the Institute of Artificial Intelligence, University of Brussels, Belgium, from 2012 to 2014. His research interests include control theory, information theory, cybersecurity, networked cyber-physical systems, multi-agent systems, and dynamical systems. He received the Best Paper Finalist Award at the International Workshop on Discrete Event Systems in 2018, and the Best Paper Award at the International Conference on Intelligent Autonomous Systems in 2014.
\end{IEEEbiographynophoto}

\vspace{-4cm}

\begin{IEEEbiographynophoto}{Mohamad Assaad} received the M.Sc.~and Ph.D.~degrees in telecommunications from Telecom ParisTech, Paris, France, in 2002 and 2006, respectively. Since 2006, he has been with the Telecommunications Department, CentraleSupelec, University of Paris-Saclay, France, where he is currently a professor. He is also a researcher at the Laboratory of Signals and Systems (CNRS), where he was the holder of the 5G Chair between 2017 and 2021. He has co-authored one book and more than 150 publications in journals and conference proceedings. He has served as a chair or a member of technical program committees of various international conferences including IEEE Wireless Communications and Networking Conference and IEEE Mobile and Wireless Networks Symposium. He has also served as an editor of several international journals including the IEEE Wireless Communications Letters and the IEEE Transactions on Network Science and Engineering. He has delivered successful tutorials on topics related to 5G systems and goal-oriented communications at different conferences. His research interests include 5G and 6G systems, fundamental networking aspects of wireless systems, semantic communications, cyber security, resource optimisation, and machine learning in wireless~networks.
\end{IEEEbiographynophoto}

\vspace{-4cm}

\begin{IEEEbiographynophoto}{John~S.~Baras} received the Diploma degree in electrical and mechanical engineering from the National Technical University of Athens, Athens, Greece, in 1970, and the M.S.~and Ph.D.~degrees in applied mathematics from Harvard University, Cambridge, MA, USA, in 1971 and 1973, respectively. He is a Distinguished University Professor and holds the Lockheed Martin Chair in Systems Engineering, with the Department of Electrical and Computer Engineering and the Institute for Systems Research (ISR), at the University of Maryland College Park. From 1985 to 1991, he was the Founding Director of the ISR. Since 1992, he has been the Director of the Maryland Center for Hybrid Networks (HYNET), which he co-founded. His research interests include systems and control, optimisation, communication networks, applied mathematics, machine learning, artificial intelligence, signal processing, robotics, computing systems, security, trust, systems biology, healthcare systems, model-based systems engineering. Dr.~Baras is a Fellow of IEEE (Life), SIAM, AAAS, NAI, IFAC, AMS, AIAA, Member of the National Academy of Inventors and a Foreign Member of the Royal Swedish Academy of Engineering Sciences. Major honours include the 1980 George Axelby Award from the IEEE Control Systems Society, the 2006 Leonard Abraham Prize from the IEEE Communications Society, the 2017 IEEE Simon Ramo Medal, the 2017 AACC Richard E. Bellman Control Heritage Award, the 2018 AIAA Aerospace Communications Award. In 2016 he was inducted in the A. J. Clark School of Engineering Innovation Hall of Fame. In 2018 he was awarded a Doctorate Honoris Causa by his alma mater the National Technical University of Athens, Greece.
\end{IEEEbiographynophoto}

\end{document}